\documentclass[12pt]{article}\usepackage{graphicx,amsfonts,amsmath,amssymb,latexsym}

\usepackage{graphicx}
\numberwithin{equation}{section}

\renewcommand{\title}[1] {%
\begingroup\begin{center}\vspace{0.0cm}\bf\Large
\addtolength{\baselineskip}{1mm} #1 \end{center}\endgroup}

\renewcommand{\author}[1] {%
\begingroup\begin{center}\vspace{0.2cm}\bf #1 \vspace{0.2cm}
\end{center}\endgroup}

\newcommand{\address}[1] {%
\begingroup\begin{center} #1 \end{center}\endgroup}

\newcommand{\addressemail}[1] {%
\begingroup\begin{center}\vspace{3mm}
\vskip-\baselineskip #1 \end{center}\endgroup}

\newcommand\ben{\begin{equation*}}
\newcommand\ebn{\end{equation*}}
\newcommand\be{\begin{equation}}
\newcommand\eb{\end{equation}}

\begin{document}
 \title{Transfer matrix eigenvectors \\
  of the Baxter-Bazhanov-Stroganov $\tau_2$-model for $N=2$}
  \vspace{0.2cm}
 \author{Oleg Lisovyy}
 \address{School of Theoretical Physics,\\
   Dublin Institute for Advanced Studies, \\
   10 Burlington Road, Dublin 4, Ireland\vspace{0.2cm}\\
   Bogolyubov Institute for Theoretical Physics \\ 14b Metrologichna
   str., 03143, Kyiv, Ukraine\vspace{0.2cm}}
 \addressemail{\texttt{olisovyy@stp.dias.ie}}
  \date{}
  \begin{abstract}
  We find a representation of the row-to-row transfer matrix of the Baxter-Bazhanov-Stroganov
  $\tau_2$-model for $N=2$ in terms of an integral over two commuting sets of grassmann variables.
  Using this representation, we explicitly calculate transfer matrix  eigenvectors and normalize them.
  It is also shown how form factors of the model can be expressed
  in terms of determinants and inverses of certain Toeplitz matrices.
  \end{abstract}
  \section{Preliminaries}
   The $\tau_2$-model was originally introduced by Baxter in the
  work  \cite{baxter1}, where it appeared in relation to the superintegrable
  case of the chiral Potts model. Later it was used by Bazhanov and Stroganov to establish
  a connection between six-vertex model and chiral Potts model
  \cite{bs}. This connection has allowed to obtain a system of functional relations
  for transfer matrices of these models \cite{baxter3} and has led
  to the derivation of exact formulas for the free energy \cite{energy} and order
  parameter \cite{order}
  of the chiral Potts model. Following the authors of \cite{bis,gps}, we will use
  instead of the name `$\tau_2$-model' the
  name `Baxter-Bazhanov-Stroganov model' (or simply `BBS model').

  BBS model is a system of spins, living on a square lattice and
  taking on $N$ values $0,1,\ldots,N-1$. The interactions exist
  only between nearest neighbours. In addition, the difference
  $b_2-b_1$ of neighbouring spins, living on the same vertical line ($b_2$ is higher than
  $b_1$) is allowed to take on
  only the values $0$ and $1$ ($\mathrm{mod}\;N$).
  Consider an elementary plaquette of the lattice, drawn
  in the Fig.~1a. Boltzmann weights $W(b_1,b_2,b_3,b_4)$, associated to this plaquette,
  are defined in the following table (our $b_1$, $b_2$, $b_3$, $b_4$ correspond to
   $d$, $a$, $b$, $c$ of \cite{baxter2} and to $b_4$, $b_1$, $b_2$, $b_3$ of \cite{bis}):

  \begin{center}
 \begin{tabular}{c|c|c}
  $b_2-b_1$ & $b_3-b_4$ & $W(b_1,b_2,b_3,b_4)$ \\
  \hline\hline
  0 & 0 & $ 1-\omega^{b_1-b_3+1}t/(yy')$ \\
  0 & 1 & $(y-\omega^{b_1-b_3+1}x')\mu'/(yy')$\\
  1 & 0 & $-(y'-\omega^{b_1-b_3+1}x)\omega\mu t/(yy')$ \\
  1 & 1 & $\;-(t-\omega^{b_1-b_3+1}xx')\omega\mu\mu'/(yy')$ \\
 \end{tabular}
 \end{center}
 Here $\omega=e^{2\pi i/N}$ and $t,x,x',y,y',\mu,\mu'$ are
 parameters. It is easily seen that the model is $\mathbb{Z}_N$-symmetric:
 if one shifts the spins $b_1,\ldots,b_4$ by $1$, all plaquette
 Boltzmann weights remain unchanged.

 \begin{figure}[h]
 \begin{center}
 \includegraphics[height=3.2cm]{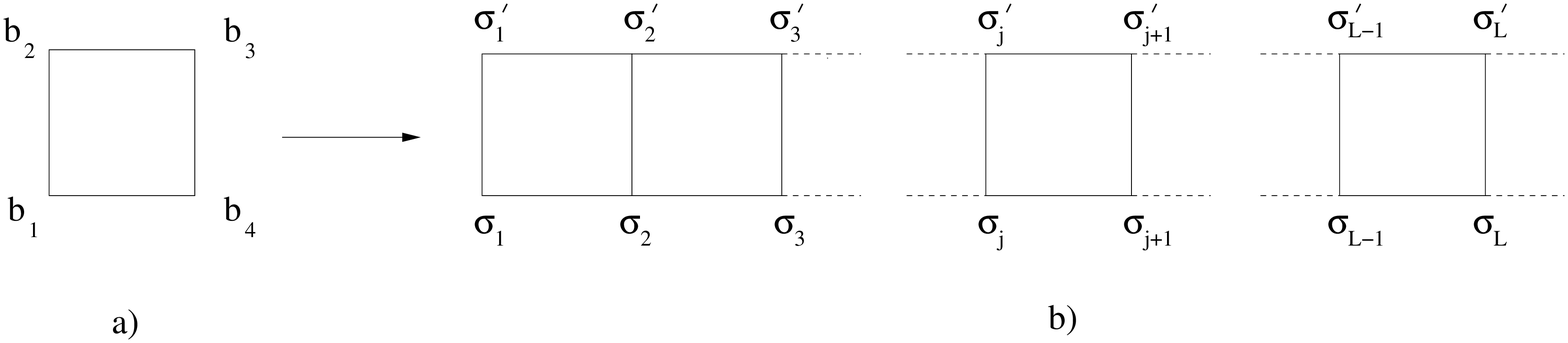}\vspace{0.2cm}\\
 \footnotesize{Fig.~1. a) numeration of spins of an elementary plaquette b)
 graphical representation of the transfer matrix}
 \end{center}
 \end{figure}

 Consider now the case $N=2$. Let us use instead of
 $b_1,\ldots,b_4$ new spin variables $\sigma_j=(-1)^{b_j}$
 ($j=1,\ldots,4$), taking on the values $\pm1$. The most general
  $\mathbb{Z}_2$-symmetric Boltzmann weight is given by the
  following formula
 \be\label{weightg}
 W(\sigma_1,\sigma_2,\sigma_3,\sigma_4)=a_0\Bigl(1+\!\!\!\sum\limits_{1\leq i<j\leq4}
 \!\!\!a_{ij}\,\sigma_i\sigma_j+a_4\,\sigma_1\sigma_2\sigma_3\sigma_4\Bigr).
 \eb
 The coefficients $a_0,\{a_{ij}\},a_4$, which correspond to BBS$_2$ model, can be written as
 \begin{eqnarray}\label{bbspars}
 a_0&=&(y+\mu t)(y'+\mu')/(4yy'), \\
 a_0a_{4}&=&(y-\mu t)(y'-\mu')/(4yy'),\\
 a_0a_{12}&=&(y-\mu t)(y'+\mu')/(4yy'), \\
 a_0a_{34}&=&(y+\mu t)(y'-\mu')/(4yy'), \\
 a_0a_{13}&=&(1+x\mu)(t+x'\mu')/(4yy'), \\
 a_0a_{24}&=&(1-x\mu)(t-x'\mu')/(4yy'), \\
 a_0a_{14}&=&(1+x\mu)(t-x'\mu')/(4yy'), \\
 \label{bbsparsf}a_0a_{23}&=&(1-x\mu)(t+x'\mu')/(4yy').
 \end{eqnarray}
 It was pointed out in \cite{bis} that these coefficients satisfy
 a `free-fermion condition'
 \be\label{ffc}
 a_4=a_{12}a_{34}-a_{13}a_{24}+a_{14}a_{23}.
 \eb

 Partition function of the model with plaquette weight
 (\ref{weightg}), satisfying the condition (\ref{ffc}), was
 calculated by Bugrij \cite{buggen} even in the case of a finite lattice.
 This result has allowed to obtain the eigenvalues of the BBS$_2$ transfer matrix
 without solving any functional relations \cite{bis}.
 From the technical point of view, the condition (\ref{ffc})
 means that the Boltzmann weight (\ref{weightg})
 can be represented as the following integral over four auxiliary grassmann variables
 $\psi^1$, $\dot{\psi}^2$, $\dot{\psi}^3$, $\psi^4$:
 \ben
 W(\sigma_1,\sigma_2,\sigma_3,\sigma_4)=\int
 d\psi^1\,d\dot{\psi}^2\,d\dot{\psi}^3\,d\psi^{4}\;\exp\Bigl\{
 a_{12}\sigma_1\sigma_2\psi^1\dot{\psi}^2-a_{13}\sigma_1\sigma_3\psi^1\dot{\psi}^3+
 a_{14}\sigma_1\sigma_4\psi^1\psi^4+\Bigr.
 \ebn
 \be\label{grweight}
 \Bigl.+a_{23}\sigma_2\sigma_3\dot{\psi}^2\dot{\psi}^3-
 a_{24}\sigma_2\sigma_4\dot{\psi}^2\psi^4+a_{34}\sigma_3\sigma_4\dot{\psi}^3\psi^4\Bigr\}\;
 e^{\psi^1}e^{\dot{\psi}^2}e^{\dot{\psi}^3}e^{\psi^4}.
 \eb
 Throughout this paper, we will use the convention that `dotted'
 grassmann variables commute with the usual ones, and that the variables
 inside each set anticommute:
 \ben
 \psi^{\alpha}\psi^{\beta}=-\psi^{\beta}\psi^{\alpha},\qquad
 \dot{\psi}^{\alpha}\dot{\psi}^{\beta}=-\dot{\psi}^{\beta}\dot{\psi}^{\alpha},\qquad
 \psi^{\alpha}\dot{\psi}^{\beta}=\dot{\psi}^{\beta}\psi^{\alpha}\qquad
 \forall\;\alpha,\beta.
 \ebn

 The method of grassmann variables was initially designed as a
 method of simple calculation of the partition function of the
 2D Ising model. It was discovered and improved by different
 authors (\cite{fradkin} is probably the earliest reference). The
 use of two commuting sets of grassmann variables, which is crucial for our
 further discussion, was suggested in \cite{bugITP}.

 The main
 drawback of the method of grassmann integration is that it does
 not give the eigenvectors of the transfer matrix, which are
 necessary ingredients in the computation of
 correlation functions and form factors. Even in the case of the
 Ising model, the only known practical way of calculation of
 these eigenvectors is the algebraic method of Kaufman
  \cite{kaufman} (it
 should be mentioned, however, that recently  a
 considerable progress has been achieved\cite{gips} in the calculation of the eigenvectors of
 the BBS transfer matrix using Sklyanin's method of separation of
 variables).
 It consists of two steps. First
 one should remark that the transfer matrix induces a rotation in a
 certain Clifford algebra. Then the eigenvectors are given by certain vectors from a Fock
 space, associated to the basis of this algebra, in which the
 above rotation is diagonal. Although Kaufman's method was later
 extended to some other free-fermion models \cite{smj5}, it does
 not seem to work neither for the general free-fermion model nor in the case of
 the BBS$_2$ model\footnote{Kaufman's method has also some drawbacks. First,
 it does not give a convenient representation of eigenvectors in terms of
 initial spin variables. Therefore, one is forced to do all the calculations
 in purely algebraic setting. Second, in the Ising case the transfer matrix
 spectrum is highly degenerate, so the eigenvectors are not determined uniquely.
 However, in the calculation of correlation functions and form factors
 one is typically interested in a very precise basis of eigenstates;
 in addition to the transfer matrix, they should also diagonalize the operator of
 discrete translations. Kaufman's method does not guarantee this last
 condition. }. The main complication, as compared to the
 Ising model case, is that one should \textit{guess} the explicit form
 of the appropriate rotation of the Clifford algebra.

 Having spent some time trying to guess the answer for the
 rotation, the author has finally found another method, which links grassmann integral
 approach with the transfer matrix formalism and allows to
 obtain the eigenvectors of the tranfer matrix of the general
 free-fermion model (i.~e. the model with plaquette weight
 (\ref{weightg}), satisfying the condition (\ref{ffc})). The
 present paper is devoted to the exposition of this method.

 This paper is organized as follows. In the next section, we find
 a convenient representation of the row-to-row transfer matrix of the periodic BBS$_2$ model
 (or, rather, general free-fermion model) in terms of a
 grassmann integral, involving two commuting sets of variables (formulas (\ref{tnsrf}),
 (\ref{ansr})). In Section~3, the eigenvectors of this transfer
 matrix are calculated (basic ansatz is given by (\ref{ansatz1})). It should be pointed out that the form of the
 answer depends on whether the number of sites in one row of the lattice is even or odd.
 In Section~4, we find a dual basis of eigenvectors and normalize
 them. It is also shown that one can express form factors of the model
 in terms of determinants and inverses of certain
 Toeplitz matrices. Finally, in the last section the above results
 are specialized to two particular cases (BBS$_2$ model and Ising model)
 and are rewritten in more common notation.

  \section{Grassmann integral representation \\ for the transfer matrix}
  Let us introduce the row-to-row transfer matrix of the
  BBS$_2$ model. It is given by the product of plaquette Boltzmann
  weights over one row (see Fig.~1b),
  \be\label{tm}
  T[\sigma,\sigma']=\prod_{j=1}^{L}
  W(\sigma_j,\sigma'_j,\sigma'_{j+1},\sigma_{j+1}),
  \eb
  where periodic boundary conditions are imposed on spin
  variables:
  \ben
  \sigma_{L+1}=\sigma_1,\qquad   \sigma'_{L+1}=\sigma'_1.
  \ebn
  This matrix naturally acts in the $2^L$-dimensional vector
  space $V$, composed of functions of $L$ spin variables
  $\sigma_1,\ldots,\sigma_L$. Namely, for any $f[\sigma]\in V$ we
  define the left action
  \ben
  (Tf)[\sigma]=\sum\limits_{[\sigma']}T[\sigma,\sigma']f[\sigma'].
  \ebn
  Partition function of the BBS$_2$ model on $L\times M$ lattice, wrapped on the
  torus, may be expressed in terms of the eigenvalues of $T$:
  \ben
  Z(L,M)=\mathrm{Tr}\;T^M=\sum\limits_{[\sigma^{(1)}]}\ldots\sum\limits_{[\sigma^{(M)}]}
  T[\sigma^{(1)},\sigma^{(2)}]\ldots
  T[\sigma^{(M)},\sigma^{(1)}].
  \ebn
  In order to compute various correlation functions, one also
  needs to know matrix elements of local field operators in
  the normalized basis of eigenstates of $T$. To obtain these
  eigenstates in an explicit form, let us first find a convenient
  representation of the transfer matrix.

  From the formulas (\ref{grweight}) and (\ref{tm}) it follows that one can write
  $T$ in the form of a grassmann integral,
  \ben
  T[\sigma,\sigma']=a_0^L\int\mathcal{D}\psi\mathcal{D}\dot{\psi}\,
  \exp\biggl\{\sum\limits_{j=1}^L\biggl(
  a_{12}\,\sigma_j\sigma'{}_{\!\! j}\,\psi_j^1\dot{\psi}_j^2-
  a_{13}\,\sigma_{j}\sigma'{}_{\!\! j+1}\,\psi_j^1\dot{\psi}_{j+1}^3+
  a_{14}\,\sigma_{j}\sigma_{j+1}\,\psi_j^1\psi_{j+1}^4+\biggr.\biggr.
  \ebn
  \ben
  \biggl.\biggl.+
  a_{23}\,\sigma'{}_{\!\! j}\sigma'{}_{\!\!
  j+1}\,\dot{\psi}_{j}^2\dot{\psi}_{j+1}^3-
  a_{24}\,\sigma'{}_{\!\!
  j}\sigma_{j+1}\,\dot{\psi}_j^2\psi_{j+1}^4+
  a_{34}\,\sigma'{}_{\!\!
  j+1}\sigma_{j+1}\,\dot{\psi}^3_{j+1}\psi^4_{j+1}
  \biggr)\biggr\}\;
  \prod\limits_{j=1}^L
  \;e^{\psi_j^1}\;e^{\dot{\psi}_j^2}\;e^{\dot{\psi}_{j+1}^3}\,e^{\psi_{j+1}^4}\,,
  \ebn
  where the measure is given by
  \ben
  \mathcal{D}\psi\,\mathcal{D}\dot{\psi}=
  \prod\limits_{j=1}^L\;\Bigl(d\psi_j^1\;d\dot{\psi}_j^2\;d\dot{\psi}_{j+1}^3\,d\psi_{j+1}^4\Bigr).
  \ebn
  Now let us make the change of integration variables:
  \ben
  \psi_j^1\rightarrow\sigma_j\,\psi_j^1,\qquad
  \dot{\psi}_j^2\rightarrow\sigma'{}_{\!\! j}\,\dot{\psi}_j^2,\qquad
  \dot{\psi}_j^3\rightarrow\sigma'{}_{\!\! j}\,\dot{\psi}_j^3,\qquad
  \psi_j^4\rightarrow\sigma_j\,\psi_j^4,\qquad\qquad j=1,\ldots,L.
  \ebn
  This change does not affect the measure, since every $\sigma_j$
  and $\sigma'{}_{\!\!j}$ appears in it twice. Spin variables
  disappear from the first (quadratic) exponential under the integral,
  but emerge in the `tail'. Namely, one
  obtains
  \be\label{tr1}
 T[\sigma,\sigma']=a_0^L\int\mathcal{D}\psi\,\mathcal{D}\dot{\psi}\;
 e^{\,S_1[\psi,\dot{\psi}]}\;
 \left[e^{\sigma_1\psi_1^1}
 \prod\limits_{j=2}^{L}\left(e^{\sigma_j\psi_j^4}e^{\sigma_j\psi_j^1}\right)
 e^{\sigma_1\psi_1^4}\right]
 \left[e^{\sigma'{}_{\!\!1}\dot{\psi}_1^2}
 \prod\limits_{j=2}^{L}\left(e^{\sigma'{}_{\!\!j}\dot{\psi}_j^3}e^{\sigma'{}_{\!\!j}\dot{\psi}_j^2}\right)
 e^{\sigma'{}_{\!\!1}\dot{\psi}_1^3}\right],
  \eb
 where $S_1[\psi,\dot{\psi}]$ can be schematically represented as
 \be\label{action1}
 S_1[\psi,\dot{\psi}]=\frac12 \left(\begin{array}{cc}
 \psi & \dot{\psi}\end{array}\right)
 \hat{D}_1
 \left(\begin{array}{c}
 \psi \\ \dot{\psi}\end{array}\right),
 \eb
 with $\left(\begin{array}{cc}\psi & \dot{\psi} \end{array}\right)=\left(\begin{array}{cccc}
 \psi^1 & \dot{\psi}^2 & \dot{\psi}^3 & \psi^4\end{array}\right)$
 and
 \be\label{dirac1}
 \hat{D}_1=\left(\begin{array}{cccc}
 0 & a_{12} & -a_{13}\,\nabla_x & a_{14}\,\nabla_x \\
 a_{12} & 0 & a_{23}\,\nabla_x & -a_{24}\,\nabla_x \\
 -a_{13}\,\nabla_{-x} & -a_{23}\,\nabla_{-x} & 0 & a_{34} \\
 -a_{14}\,\nabla_{-x} & -a_{24}\,\nabla_{-x} & a_{34} & 0
 \end{array}\right).
 \eb
 Here $\nabla_x$ denotes the operator, shifting the lower indices of grassmann variables
 by 1. It obeys periodic boundary condition $\left(\nabla_x\right)^L=1$. For example,
 one has
 \ben
 \psi^1\,\nabla_{x}\,\dot{\psi}^{\,3}=\sum\limits_{j=1}^{L-1}\,\psi^1_j\,\dot{\psi}^3_{j+1}
 +\psi^1_L\,\dot{\psi}^3_{1}\,.
 \ebn

 Note that in (\ref{tr1}) we have rearranged the tail, assembling
 together the exponentials, containing the same spin variables. It
 seems, however, that the exponentials $e^{\sigma'{}_{\!\!1}\dot{\psi}_1^3}$
 and $e^{\sigma_1\psi_1^4}$ are not on their `right' places. One
 may correct this, observing that for any function $F[\psi]$ and for any grassmann variable
 $\psi_{\alpha}$ we have the identity
 \ben
 F[\psi]\,e^{\psi_{\alpha}}=e^{\psi_{\alpha}}\frac{F[\psi]+F[-\psi]}{2}+
 e^{-\psi_{\alpha}}\frac{F[\psi]-F[-\psi]}{2}\,.
 \ebn
 Now, introducing the notation
 \ben
 F_1[\psi]=e^{\sigma_1\psi_1^1}
 \prod\limits_{j=2}^{L}\left(e^{\sigma_j\psi_j^4}e^{\sigma_j\psi_j^1}\right),\qquad
 F_2[\dot{\psi}]=e^{\sigma'{}_{\!\!1}\dot{\psi}_1^2}
 \prod\limits_{j=2}^{L}\left(e^{\sigma'{}_{\!\!j}\dot{\psi}_j^3}
 e^{\sigma'{}_{\!\!j}\dot{\psi}_j^2}\right),
 \ebn
 one may rewrite the tail as
 \begin{eqnarray}
 \label{pull}F_1[\psi]\,e^{\sigma_1\psi_1^4}F_2[\dot{\psi}]\,e^{\sigma'{}_{\!\!1}\dot{\psi}_1^3}=
 \biggl\{e^{\sigma_1\psi_1^4}\frac{F_1[\psi]+F_1[-\psi]}{2}+
 e^{-\sigma_1\psi_1^4}\frac{F_1[\psi]-F_1[-\psi]}{2}\biggr\}\times\\
 \nonumber\times\biggl\{e^{\sigma'{}_{\!\!1}\dot{\psi}_1^3}\frac{F_2[\dot{\psi}]+F_2[-\dot{\psi}]}{2}+
 e^{-\sigma'{}_{\!\!1}\dot{\psi}_1^3}\frac{F_2[\dot{\psi}]-F_2[-\dot{\psi}]}{2}\biggr\}.\quad
 \end{eqnarray}
 Expanding this last expression, one obtains 16 terms. However,
 some of these terms are equivalent, since the simultaneous change
 of the signs of all $\psi$ and $\dot{\psi}$ does not affect
 the value of the integral. Then one may easily check that
 (\ref{pull}) may be replaced (after appropriate change of
 variables) by the following combination, containing only 4 terms:
 \begin{eqnarray}
 \label{4terms}
 F_1[\psi]\,e^{\sigma_1\psi_1^4}F_2[\dot{\psi}]\,e^{\sigma'{}_{\!\!1}\dot{\psi}_1^3}
 \rightarrow\frac12\left\{e^{-\sigma_1\psi_1^4}F_1[\psi]\,
 e^{-\sigma'{}_{\!\!1}\dot{\psi}_1^3}F_2[\dot{\psi}]+
 e^{\sigma_1\psi_1^4}F_1[-\psi]\,
 e^{-\sigma'{}_{\!\!1}\dot{\psi}_1^3}F_2[\dot{\psi}]+\right.\\
 \nonumber\left. +e^{\sigma_1\psi_1^4}F_1[\psi]\,
 e^{\sigma'{}_{\!\!1}\dot{\psi}_1^3}F_2[\dot{\psi}]-
 e^{-\sigma_1\psi_1^4}F_1[-\psi]\,
 e^{\sigma'{}_{\!\!1}\dot{\psi}_1^3}F_2[\dot{\psi}]\right\}.
 \end{eqnarray}

 The third term of the last expression has the desired form and there is no need to
 transform it further. If we make the substitution $\dot{\psi}_1^3\rightarrow -\dot{\psi}_1^3$,
 $\psi_1^4\rightarrow -\psi_1^4$  in the
 integral, corresponding to the first term, it will have almost
 the same structure. The only difference is that the boundary
 condition for the shift operator~$\nabla_x$ becomes antiperiodic:
 $\left(\nabla_x\right)^L=-1$.

 Next one should remark that the second and the fourth term in (\ref{4terms}) can
 be obtained from the first and the third one, respectively, by
 changing the signs of the spins $\sigma_1,\ldots,\sigma_L$. This
 change can be realized, using the operator of spin reflection
 $U$, whose defining property is that $(Uf)[\sigma]=f[-\sigma]$ for any vector
 $f[\sigma]\in V$. Matrix elements of $U$ may be explicitly
 written as
 \ben
 U[\sigma,\sigma']=\prod\limits_{j=1}^L\frac{1-\sigma_j\,\sigma'{}_{\!\!j}}{2}\,.
 \ebn

 Summarizing the above observations, we obtain the following
 representation for the transfer matrix:
 \be\label{tstr}
 T=\frac{1+U}{2}\,T^{NS}+\frac{1-U}{2}\,T^{R},
 \eb
 where
 \be\label{tnsr1}
 T^{NS(R)}[\sigma,\sigma']=a_0^L\int\mathcal{D}\psi\,\mathcal{D}\dot{\psi}\;
 \exp\left\{\,S_1^{NS(R)}[\psi,\dot{\psi}]\right\}\;
 \prod\limits_{j=1}^{L}\left(e^{\sigma_j\psi_j^4}e^{\sigma_j\psi_j^1}\right)
 \prod\limits_{j=1}^{L}\left(e^{\sigma'{}_{\!\!j}\dot{\psi}_j^3}
 e^{\sigma'{}_{\!\!j}\dot{\psi}_j^2}\right),
 \eb
 and both actions $S_1^{NS(R)}[\psi,\dot{\psi}]$ are defined by the
 formulas (\ref{action1})--(\ref{dirac1}). Upper indices NS and R
 correspond to antiperiodic (Neveu-Schwartz) and periodic (Ramond)
 boundary conditions, satisfied by the shift operator~$\nabla_x$.

 The matrices $P_{\pm}=\frac{1\pm U}{2}$ have the properties of
 projectors, i.~e. $P_{\pm}^2=P_{\pm}$; thus their eigenvalues are
 equal to either 0 or 1. The eigenvectors, corresponding to zero
 eigenvalues of $P_+$ ($P_-$), are odd (even) under
 spin reflection. It means that a vector $f[\sigma]\in V$ will
 satisfy $(P_+ f)[\sigma]=0$ ($(P_- f)[\sigma]=0$) iff $f[\sigma]=-f[-\sigma]$
 (respectively, $f[\sigma]=f[-\sigma]$). Analogously, the eigenvectors of $P_+$ ($P_-$)
 with eigenvalue 1 are even (odd) under spin reflection.

 The operator $U$ commutes with the transfer matrix $T$. Therefore, these two matrices can
 be diagonalized simultaneously and one may choose the
 eigenvectors of $T$ so that they are either even or odd under
 the action of $U$. Let us take an even eigenvector $f_{e}$ of
 $T$, and denote by $\lambda_{f_e}$ the corresponding eigenvalue.
 Acting on $f_{e}$ by both sides of the
 relation (\ref{tstr}), and using the fact that $U$ commutes with
 $T^{NS}$ and $T^R$ as well, one obtains
 \ben
 \lambda_{f_e} f_{e} = Tf_{e}=
 (T^{NS}P_++T^{R}P_-)f_{e}=
 T^{NS}f_{e},
 \ebn
 that is, any even eigenvector of $T$ is an eigenvector of $T^{NS}$
 with the same eigenvalue. Similarly, any odd eigenvector of $T$
 is an eigenvector of $T^{R}$. Conversely, any even eigenvector of $T^{NS}$ and any odd
 eigenvector of $T^R$ are eigenvectors of $T$. Therefore,
 the set of all transfer matrix eigenstates splits into two parts:
 NS-sector (even eigenvectors of $T^{NS}$) and R-sector (odd
 eigenvectors of $T^R$). The problem of diagonalization of $T$ is
 then reduced to the calculation of eigenvectors and eigenvalues
 of matrices $T^{NS}$ and $T^R$, given by the formula
 (\ref{tnsr1}).\vspace{0.2cm}\\
 \textbf{Remark}. Having diagonalized $T^{NS}$ and $T^R$, one also
 gets for free the solution of the BBS$_2$ model with antiperiodic
 boundary conditions for spin variables (in one direction). It is easy to understand
 that the transfer matrix of such model is given by
 \ben
 T^a=\frac{1-U}{2}\,T^{NS}+\frac{1+U}{2}\,T^{R},
 \ebn
 and, therefore, the set of its eigenstates is composed of odd
 eigenvectors of $T^{NS}$ and even eigenvectors of
 $T^R$.\vspace{0.1cm}

 The representation (\ref{tnsr1}) can be simplified even further by
 integrating over fermionic degrees of freedom, which
 are not coupled to spin variables. We mean the following:
 elementary factors from the products of (\ref{tnsr1}) can be
 written as
 \ben
 e^{\sigma_j\psi_j^4}e^{\sigma_j\psi_j^1}=e^{-\psi_j^1\,\psi_j^4}\,
 e^{\,\sigma_j\left(\psi_j^1+\psi_j^4\right)},\qquad\qquad
 e^{\sigma'{}_{\!\!j}\dot{\psi}_j^3}
 e^{\sigma'{}_{\!\!j}\dot{\psi}_j^2}=e^{-\dot{\psi}_j^2\,\dot{\psi}_j^3}\,
 e^{\,\sigma'{}_{\!\!j}\left(\dot{\psi}_j^2+\dot{\psi}_j^3\right)}.
 \ebn
 Let us now introduce instead of $\psi$ and $\dot{\psi}$ new
 grassmann variables
 \be\label{newv1}
 \varphi_j=\psi^1_j+\psi^4_j,\qquad
 \dot{\varphi}_j=\dot{\psi}^2_j+\dot{\psi}^3_j,\qquad
 \eta_j=\psi^4_j,\qquad \dot{\eta}_j=\dot{\psi}^3_j,\qquad\qquad
 j=1,\ldots,L.
 \eb
 Since the jacobian of the transformation (\ref{newv1}) is equal to 1, the
 integration measure transforms as
 \ben
 \mathcal{D}\psi\,\mathcal{D}\dot{\psi}\rightarrow
 \mathcal{D}\varphi\,\mathcal{D}\dot{\varphi}\,\mathcal{D}\eta\,\mathcal{D}\dot{\eta}=
 \prod\limits_{j=1}^L \Bigl(d\varphi_j\,d\dot{\varphi}_j\,d\eta_j\,d\dot{\eta}_j\,\Bigr).
 \ebn
 Then the integral (\ref{tnsr1}) may be rewritten as
 \be\label{tnsr2}
 T^{NS(R)}[\sigma,\sigma']=
 a_0^L\int \mathcal{D}\varphi\,\mathcal{D}\dot{\varphi}\,\mathcal{D}\eta\,\mathcal{D}\dot{\eta}\;
 \exp\left\{\,S_2^{NS(R)}[\varphi,\dot{\varphi},\eta,\dot{\eta}]\right\}\;
 \prod\limits_{j=1}^{L}e^{\sigma_j\varphi_j}
 \prod\limits_{j=1}^{L}e^{\sigma'{}_{\!\!j}\dot{\varphi}_j},
 \eb
 where the action
 $S_2^{NS(R)}[\varphi,\dot{\varphi},\eta,\dot{\eta}]$ is given by
 \ben
 S_2^{NS(R)}[\varphi,\dot{\varphi},\eta,\dot{\eta}]=\frac12\,
 \left(\begin{array}{cccc}\varphi & \dot{\varphi} & \dot{\eta} & \eta
 \end{array}\right)\hat{D}_2
 \left(\begin{array}{cccc}\varphi & \dot{\varphi} & \dot{\eta} & \eta
 \end{array}\right)^T,
 \ebn
 \small
 \ben
 \hat{D}_2=\left(\begin{array}{cccc}
 0 & \!\!a_{12} & -a_{12}-a_{13}\,\nabla_x & \!\!\!\!\!\!-1+a_{14}\,\nabla_x \\
 a_{12} & \!\!0 & -1+a_{23}\,\nabla_x & \!\!\!\!\!\!-a_{12}-a_{24}\,\nabla_x \\
 -a_{12}-a_{13}\,\nabla_{-x} & \!\!1-a_{23}\,\nabla_{-x} &
 -a_{23}(\nabla_x-\nabla_{-x}) & \!\!\!\!\!\!a_{12}+a_{34}+a_{13}\nabla_{-x}+a_{24}\nabla_x \\
 1-a_{14}\,\nabla_{-x} & \!\!-a_{12}-a_{24}\,\nabla_{-x} &
 a_{12}+a_{34}+a_{13}\nabla_x+a_{24}\nabla_{-x} & \!\!\!\!\!\!-a_{14}(\nabla_x-\nabla_{-x})
 \end{array}\right).
 \ebn
 \normalsize
 Let us now integrate over $\eta$ and $\dot{\eta}$ in the
 representation (\ref{tnsr2}). This integration can be done
 relatively easily, since $S_2^{NS(R)}[\varphi,\dot{\varphi},\eta,\dot{\eta}]$
 is diagonalized by Fourier transformation. Namely, if one denotes
 \be\label{fourier}
 \left(\begin{array}{cccc}\varphi_p & \dot{\varphi}_p & \dot{\eta}_p &
 \eta_p \end{array}\right)=\frac{1}{\sqrt{L}}\sum\limits_{j=1}^L
 e^{-i p\, j}\left(\begin{array}{cccc}\varphi_j & \dot{\varphi}_j & \dot{\eta}_j &
 \eta_j \end{array}\right),
 \eb
 then
 \be\label{sumnsr1}
 S_2^{NS(R)}[\varphi,\dot{\varphi},\eta,\dot{\eta}]=\frac12\;{\sum\limits_{p}}^{NS(R)}
 \left(\begin{array}{cccc}\varphi_{-p} & \dot{\varphi}_{-p} & \dot{\eta}_{-p} &
 \eta_{-p} \end{array}\right)\hat{D}_2(p)
 \left(\begin{array}{cccc}\varphi_p & \dot{\varphi}_p & \dot{\eta}_p &
 \eta_p \end{array}\right)^T,
 \eb
 where the one-mode matrix $\hat{D}_2(p)$ is given by
 \ben
 \hat{D}_2(p)=
 \left(\begin{array}{cccc}
 0 & \!\!\!\!a_{12} & \!-a_{12}-a_{13}\,e^{ip} & \!\!\!\!\!\!-1+a_{14}\,e^{ip} \\
 a_{12} & \!\!\!\!0 & \!-1+a_{23}\,e^{ip} & \!\!\!\!\!\!-a_{12}-a_{24}\,e^{ip} \\
 -a_{12}-a_{13}\,e^{-ip} & \!\!\!\!1-a_{23}\,e^{-ip} &
 \!-2i\,a_{23}\sin p & \!\!\!\!\!\!a_{12}+a_{34}+a_{13}e^{-ip}+a_{24}e^{ip} \\
 1-a_{14}\,e^{-ip} & \!\!\!\!-a_{12}-a_{24}\,e^{-ip} &
 \! a_{12}+a_{34}+a_{13}e^{ip}+a_{24}e^{-ip} &
 \!\!\!\!\!\!-2i\,a_{14}\sin p
 \end{array}\right)
 \ebn
 and the indices NS and R in the sum (\ref{sumnsr1}) mean that the corresponding
 quasimomenta run over Neveu-Schwartz values
 ($p=\frac{2\pi}{L}\left(j+\frac12\right)$, $j=0,1,\ldots,L-1$)
 or, correspondingly, Ramond values ($p=\frac{2\pi}{L}\,j$,
 $j=0,1,\ldots,L-1$). Note also that the integration measure can
 be written as
 \ben
 \mathcal{D}\varphi\,\mathcal{D}\dot{\varphi}\,\mathcal{D}\eta\,\mathcal{D}\dot{\eta}=
 {\prod\limits_{p}}^{NS(R)} \Bigl(d\varphi_p\,d\dot{\varphi}_p\,d\eta_p\,d\dot{\eta}_p\,\Bigr)
 \ebn

  Thus the $2L$-fold integral over $\eta$ and $\dot{\eta}$ in the
  representation (\ref{tnsr2}) factorizes into a product of 4-fold
  (over $\eta_{\pm p}$, $\dot{\eta}_{\pm p}$) and 2-fold integrals.
  Double integrals correspond to the mode $p=0$
  (always present in the Ramond sector) and $p=\pi$ (present in the Ramond sector
  for even $L$ and in the Neveu-Schwartz sector for odd $L$). After a little bit
  cumbersome but nevertheless straightforward calculation one then
  obtains
  \be\label{tnsrf}
 T^{NS(R)}[\sigma,\sigma']=
 \zeta^{NS(R)}\int \mathcal{D}^{NS(R)}\varphi\,\mathcal{D}^{NS(R)}\dot{\varphi}\;
 \exp\left\{\,S^{NS(R)}[\varphi,\dot{\varphi}]\right\}\;
 \prod\limits_{j=1}^{L}e^{\sigma_j\varphi_j}
 \prod\limits_{j=1}^{L}e^{\sigma'{}_{\!\!j}\dot{\varphi}_j},
  \eb
  where
  \ben
   \mathcal{D}^{NS(R)}\varphi={\prod\limits_{p}}^{NS(R)}d\varphi_p\,,\qquad
    \mathcal{D}^{NS(R)}\dot{\varphi}={\prod\limits_{p}}^{NS(R)}d\dot{\varphi}_p\,,
  \ebn
  \be\label{zetans}
  \zeta^{NS(R)}=a_0^L\;{\prod\limits_{p}}^{NS(R)}\chi^{1/2}_p,
  \eb
  \ben
  \chi_p=\left[a_{12}+a_{34}+(a_{13}+a_{24})\cos p\right]^2+
  \left[(a_{13}-a_{24})^2+4a_{14}a_{23}\right]\sin^2 p,
  \ebn
  and the action $S^{NS(R)}[\varphi,\dot{\varphi}]$ is given by
  \be\label{ansr}
 S^{NS(R)}[\varphi,\dot{\varphi}]=\frac12\;{\sum\limits_{p}}^{NS(R)}
 \left(\begin{array}{cc} \varphi_{-p} &
 \dot{\varphi}_{-p}\end{array}\right)
 \left(\begin{array}{cc} G_{11}(p) & G_{12}(p) \\ G_{21}(p) & G_{22}(p) \end{array}\right)
 \left(\begin{array}{c}\varphi_{p} \\
 \dot{\varphi}_{p}\end{array}\right),
  \eb
  with
  \ben
  \chi_p\,G_{11}(p)=2i \sin p\,\Bigl[
  a_{23}+a_{12}a_{24}+a_{13}a_{34}+a_{14}a_{4}-2(a_{14}a_{23}-a_{13}a_{24})\cos
  p\Bigr],
  \ebn
  \ben
  \chi_p\,G_{22}(p)=2i\sin p\,\Bigl[
  a_{14}+a_{12}a_{13}+a_{34}a_{24}+a_{23}a_4-2(a_{14}a_{23}-a_{13}a_{24})\cos
  p  \Bigr],
  \ebn
  \begin{eqnarray*}
  \chi_p\,G_{12}(p)=\chi_p\,G_{21}(-p)&=&
  \Bigl[(a_{12}+a_{34})(a_4+1)-(a_{14}+a_{23})(a_{13}+a_{24})\Bigr]+\\
  &+&\Bigl[(a_{13}+a_{24})(a_4+1)-(a_{14}+a_{23})(a_{12}+a_{34})\Bigr]\cos
  p\;+\\
  &+&\Bigl[(a_{24}-a_{13})(a_4-1)+(a_{14}-a_{23})(a_{12}-a_{34})\Bigr]
  i\sin p\;.
  \end{eqnarray*}
  As we will see in the next section, this final representation
  for $T^{NS(R)}$ (given by the formulas (\ref{tnsrf}),
  (\ref{ansr})) allows to obtain all transfer matrix eigenvectors almost
  immediately. Concrete form of the functions $G_{ij}$ ($i,j=1,2$)
  does not play any essential role.

  For further convenience and making parallels with the work \cite{bis},
  let us also introduce the notation $v_p=4\chi_p\,G_{12}(p)$ and
  \ben
  u_p=2\chi_p\Bigl(1-G_{11}(p)G_{22}(p)+G_{12}(p)G_{21}(p)\Bigr)=
  \ebn
  \ben
  =2\,\Bigl[(1+a_4)^2+(a_{12}+a_{34})^2+(a_{13}+a_{24})^2+(a_{14}+a_{23})^2\Bigr]
  +4\,\Bigl[(a_{12}+a_{34})(a_{13}+a_{24})-(a_{14}+a_{23})(1+a_4)\Bigr]\cos
  p\,.
  \ebn
  \textbf{Remark}. It should be pointed out
  that we did not care about the correct overall sign of $T^{NS(R)}$ in the
  representation (\ref{tnsrf}). However, using the fact that for
  $a_{12}=a_{13}=a_{14}=a_{23}=a_{24}=a_{34}=0$ all the
  eigenvalues of $T$ should be equal to $2^La_0^L$, one can restore this sign at any stage.

  \section{Transfer matrix eigenvectors}
  It appears that the form of the eigenvectors of $T^{NS}$ and $T^R$ depends on whether $L$ is even or odd.
  Moreover, quasiparticle interpretation of the eigenvectors and eigenvalues is different
  in different regions of parameters of the BBS$_2$ model. Below we
  will consider various cases in order of increasing difficulty.
  \subsection{NS-sector, even $L$}
  If $L$ is even, then the Neveu-Schwartz spectrum of quasimomenta
  does not contain the values 0 and $\pi$ (the only values with
  the property  $p=-p\;\mathrm{mod}\;2\pi$).

  The simplest ansatz for an eigenvector $f[\sigma]\in V$ of the
  matrix $T^{NS}$ is given by an integral over $L$ auxiliary
  grassmann  variables $\xi_1,\ldots,\xi_L$:
  \be\label{ansatz0}
  f[\sigma]=\int\mathcal{D}^{NS}\xi\;
  \exp\left\{{\sum\limits_{p}}^{\frac{NS}{2}}\xi_{-p}\,A(p)\,\xi_p\right\}
  \prod\limits_{j=1}^L
  e^{\sigma_j\xi_j}.
  \eb
  Here $A(p)=-A(-p)$ is an unknown odd function to be determined, and $\{\xi_p\}$
  denote  Fourier components of $\xi$. The indices $\frac{NS}{2}$ and
  $\frac{R}{2}$ in sums and products will be used to indicate that
  the corresponding operations involve only those Neveu-Schwartz
  and Ramond quasimomenta, which lie in the \textit{open} interval
  $(0,\pi)$ (for the NS-sector and even $L$, this is exactly one half of the Brillouin
  zone). Note that for even $L$ the function (\ref{ansatz0}) is even
  with respect to  the action of
  $U$: the reversal of all spins is
  equivalent to the change of variables $\xi\rightarrow -\xi$.

  Let us now act on $f[\sigma]$ by the matrix $T^{NS}$. Since the
  fields $\dot{\varphi}$ and $\xi$ commute, the sum
  $\sum\limits_{[\sigma']}T^{NS}[\sigma,\sigma']f[\sigma']$ can be
  easily evaluated and one obtains
  \ben
  (T^{NS}f)[\sigma]=2^L\,\zeta^{NS}\int\mathcal{D}^{NS}\varphi\,\mathcal{D}^{NS}\dot{\varphi}\;\;
  e^{\,S^{NS}[\varphi,\dot{\varphi}]}\prod\limits_{j=1}^L
  e^{\sigma_j\varphi_j}\;\times
  \ebn \ben\times\int\mathcal{D}^{NS}\xi\;\exp\left\{
  {\sum\limits_{p}}^{\frac{NS}{2}}\xi_{-p}\,A(p)\,\xi_p+
  \sum\limits_{j=1}^L\dot{\varphi}_j\,\xi_j\right\}.
  \ebn
  After integration over $\xi$ one finds the exponential of a
  quadratic form in $\dot{\varphi}$,
  \ben
  \int\mathcal{D}^{NS}\xi\,\exp\left\{
  {\sum\limits_{p}}^{\frac{NS}{2}}\xi_{-p}\,A(p)\,\xi_p+
  \sum\limits_{j=1}^L\dot{\varphi}_j\,\xi_j\right\}=\ebn
  \ben
  =\int\mathcal{D}^{NS}\xi\,\exp\left\{
  {\sum\limits_{p}}^{\frac{NS}{2}}\Bigl(\xi_{-p}\,A(p)\,\xi_p+
  \dot{\varphi}_{-p}\,\xi_p+\dot{\varphi}_{p}\,\xi_{-p}\Bigr)\right\}=
  \ebn
  \ben
  =\left({\prod\limits_{p}}^{\frac{NS}{2}} A(p)\right)\,\exp\left\{
  -{\sum\limits_{p}}^{\frac{NS}{2}}\dot{\varphi}_{-p}\,A^{-1}(p)\,\dot{\varphi}_p\right\},
  \ebn
  which can then be pulled through the `linear' exponentials. Then one may integrate
  over $\dot{\varphi}$ and obtain
  \ben
  (T^{NS}f)[\sigma]=2^L\,\zeta^{NS}\,{\prod\limits_{p}}^{\frac{NS}{2}}\Bigl(1-A(p)\,G_{22}(p)\Bigr)
  \int\mathcal{D}^{NS}\varphi\;\exp\left\{
  {\sum\limits_{p}}^{\frac{NS}{2}}\varphi_{-p}\,A'(p)\, \varphi_p\right\}
  \prod\limits_{j=1}^L
  e^{\,\sigma_j\varphi_j},
  \ebn
  with
  \ben
  A'(p)=G_{11}(p)+\frac{A(p)G_{12}(p)G_{21}(p)}{1-A(p)G_{22}(p)}\,.
  \ebn
  Therefore, the function (\ref{ansatz0}) will be an eigenvector of $T^{NS}$ iff
  for all NS-values of $p$ from the interval $(0,\pi)$ one has
  $A(p)=A'(p)$.
  This equation is quadratic in
  $A(p)$, and its roots are given by
  \be\label{1mroots}
  A^{\pm}(p)=\frac{1+G(p)
  \mp\sqrt{\bigl(1-G(p)\bigr)^2-4G_{12}(p)G_{21}(p)}}{2\,G_{22}(p)}\,,
  \eb
  where we have introduced the notation
  \ben
  G(p)=G_{11}(p)G_{22}(p)-G_{12}(p)G_{21}(p)\,.
  \ebn
  Thus the formula (\ref{ansatz0}) gives $2^{L/2}$ eigenvectors of $T^{NS}$, corresponding
  to different choices of the set of one-mode roots.

  One can take, for instance, $A(p)=A^+(p)$ for
  all $p\in(0,\pi)$. The vector, corresponding to this particular choice, will be
  denoted by $|vac\rangle_{NS}$, since under some conditions, satisfied by
  the parameters of the BBS$_2$ model, it corresponds to the
  eigenvalue with maximum modulus. Similarly, if we choose
  $A(p)=A^-(p)$ for some values $p_1,\ldots,p_k\in(0,\pi)$, and $A(p)=A^+(p)$ for all the
  other NS-quasimomenta from the interval $(0,\pi)$, then
  the corresponding eigenvector will be denoted by $|p_1,-p_1;\ldots
  p_k,-p_k\rangle_{NS}$. The origin of this notation will become
  clear soon.

  In order to find all the eigenvectors of $T^{NS}$, only a slight
  generalization of the ansatz (\ref{ansatz0}) is needed. Namely,
  let us define
  \be\label{ansatz1}
  f^{NS}_{\{i_p\}}[\sigma]=\int\mathcal{D}^{NS}\xi\;\;{\prod\limits_{p}}^{\frac{NS}{2}}
  F_{i_p}(\xi_{-p},\xi_p)\;{\prod\limits_{j=1}^L}\;e^{\,\sigma_j\xi_j}.
  \eb
  Here each of the indices $\{i_p\}$ can take any of the four values,
  which we will conventionally denote by 1, 2, 3, and 4. Corresponding
  functions $F_i(\xi_{-p},\xi_p)$ are defined as follows:
  \begin{eqnarray}\label{efs}
  F_1(\xi_{-p},\xi_p)&=&\exp\Bigl(\xi_{-p}\,A^+(p)\,\xi_p\Bigr),\\
  F_2(\xi_{-p},\xi_p)&=&\xi_{-p}\,,\\
  F_3(\xi_{-p},\xi_p)&=&\xi_{p}\,,\\
  \label{endefs}F_4(\xi_{-p},\xi_p)&=&\exp\Bigl(\xi_{-p}\,A^-(p)\,\xi_p\Bigr).
  \end{eqnarray}
  Similarly to the above, one should act on $f^{NS}_{\{i_p\}}[\sigma]$
  by $T^{NS}$, then to sum over the intermediate spin variables, to
  integrate the result over $\xi$ and, finally, over
  $\dot{\varphi}$. Then it is straightforward to verify that the formulas
  (\ref{ansatz1})--(\ref{endefs}) indeed define an eigenvector of
  $T^{NS}$ with the eigenvalue
  \be\label{eigenvalue}
  \Lambda^{NS}_{\{i_p\}}=2^L\,\zeta^{NS}\,{\prod\limits_{p}}^{\frac{NS}{2}}\lambda_{i_p}(p),
  \eb
 where  `one-mode' eigenvalues are given by
  \begin{eqnarray}\label{lambdas}
  \lambda_1(p)&=&1-A^+(p)G_{22}(p),\\
  \label{lambdas02}\lambda_2(p)&=&G_{12}(p),\\
  \label{lambdas03}\lambda_3(p)&=&G_{21}(p),\\
  \label{lambdas04}\lambda_4(p)&=&1-A^-(p)G_{22}(p).
  \end{eqnarray}
  One can also rewrite them in the following way (see the end of the previous
  section for the notations):
  \be\label{lambdas2}
  \lambda_1(p)=\frac{u_p+\sqrt{u_p^2-v_p
  v_{-p}}}{4\chi_p},\quad
  \lambda_2(p)=\frac{v_p}{4\chi_p},\quad
  \lambda_3(p)=\frac{v_{-p}}{4\chi_p},\quad
  \lambda_4(p)=\frac{u_p-\sqrt{u_p^2-v_p v_{-p}}}{4\chi_p}\,.
  \eb
  The total number of found eigenstates is equal to $4^{L/2}=2^L$,
  and thus the diagonalization of the matrix $T^{NS}$ is completed.

  Let us now turn to quasiparticle interpretation of eigenvalues
  and eigenvectors. It follows from (\ref{zetans}),
  (\ref{eigenvalue}), (\ref{lambdas2}) that the eigenvalues can be
  written in the form
  \be\label{eigenphys1}
  \Lambda_{\{i_p\}}^{NS}=\Lambda^{NS}_{max}\,
  {\prod\limits_{\;\;p|i_p=2}}^{\!\!\!\frac{NS}{2}}\;\frac{v_p}{\rho_p}
  \;{\prod\limits_{\;\;p|i_p=3}}^{\!\!\!\frac{NS}{2}}\;\frac{v_{-p}}{\rho_p}
  \;{\prod\limits_{\;\;p|i_p=4}}^{\!\!\!\frac{NS}{2}}\;\frac{v_p\,v_{-p}}{\rho^{\,2}_p}\,,
  \eb
  where
  \be\label{lnsm}
  \Lambda^{NS}_{max}=a_0^L\;{\prod\limits_p}^{NS}\rho_p^{\,1/2},\qquad
  \rho_p=u_p+\sqrt{u_p^2-v_pv_{-p}}\;.
  \eb
  \textbf{Remark}. The expression $u_p^2-v_pv_{-p}$ is a quadratic polynomial in $\cos p$.
  For the sake of simplicity, it will be assumed
  that the parameters $\left\{a_{ij}\right\}_{1\leq i<j\leq 4}$
  are all real and chosen so that this polynomial has no roots
  inside the interval $(-1,1)$ (for example, this condition is satisfied, if one takes
  $a_{12}=a_{34}$ and $a_{13}=a_{24}$).
  It means, in particular, that $u_p^2-v_pv_{-p}$ is non-negative
  and  has local extrema only at the points $p=0$
  and $p=\pi$. \vspace{0.2cm}

  Consider also the operator of translations in discrete space $R$. Its
  action on an arbitrary vector $f[\sigma]\in V$ is defined as
  \be\label{translation}
  \bigl(Rf\bigr)(\sigma_1,\sigma_2,\ldots,\sigma_L)=f(\sigma_2,\sigma_3,\ldots,\sigma_1).
  \eb
  This operator commutes with the transfer matrix $T$, with
  the matrices $T^{NS}$ and $T^R$, and also with the operator $U$ of
  spin reflection. Since we have already diagonalized $T^{NS}$ and
  obtained nondegenerate spectrum, the eigenvectors
  (\ref{ansatz1}) should diagonalize $R$ as well. Actually, it is
  not difficult to verify that
  \ben
  (Rf^{NS}_{\{i_p\}})[\sigma]={\prod\limits_{\;\;p|i_p=2}}^{\!\!\!\frac{NS}{2}}\;e^{ip}\;
  {\prod\limits_{\;\;p|i_p=3}}^{\!\!\!\frac{NS}{2}}\;e^{-ip}\;\;f^{NS}_{\{i_p\}}[\sigma].
  \ebn

  Now it is clear that the eigenvectors of $T^{NS}$ can be labelled by the collections
  of distinct NS-quasimomenta and interpreted as multiparticle
  states. One-particle  energy is given by
  \ben
  \varepsilon(p)=-\ln\frac{v_p}{\rho_p}.
  \ebn
  It may have a non-zero imaginary part, which is a general
  consequence of the fact that the transfer matrix $T$ of the
  BBS$_2$ model is not symmetric.
  The eigenstate, which contains particles
  with the momenta $p_1,\ldots,p_k$, will be denoted by $|p_1,\ldots,p_k\rangle_{NS}$.
  In order to determine, which one of the functions
  (\ref{ansatz1}) gives the explicit form of this vector, one
  should decompose the set of momenta of particles from the state $|p_1,\ldots,p_k\rangle_{NS}$
  into three parts:
  pairs of the form $\pm p_j$, `unpaired' momenta from the interval $(0,\pi)$,
  and `unpaired' momenta from the interval $(\pi,2\pi)$.
  Then in the ansatz (\ref{ansatz1}) one should set
  \begin{itemize}
  \item $i_p=4$, if $\pm p$ appears in the first part,
  \item $i_p=2$, if $p$ appears in the second part,
  \item $i_p=3$, if $-p$ appears in the third part,
  \item $i_p=1$ for all the other values of $p$.
  \end{itemize}
  This procedure establishes the correspondence between the
  formulas (\ref{ansatz1}) and usual quasiparticle
  notation.\vspace{0.2cm}\\
  \textbf{Remark}.
  Recall that only even eigenvectors of $T^{NS}$ diagonalize the
  full transfer matrix $T$ as well. It means that the total number of
  appearances of $i_p=2$ and $i_p=3$ in the functions
  (\ref{ansatz1}) should be even. In other words,
  NS-eigenstates of $T$  should contain even number of particles.

  \subsection{NS-sector, odd $L$}
  When $L$ is odd, the Neveu-Schwartz spectrum of quasimomenta
  contains the value $p=\pi$. To take into account this special
  mode, it is sufficient to slightly modify the ansatz
  (\ref{ansatz1}). Let us consider
  \be\label{ansatz2}
  f^{NS}_{\{i_p\}}[\sigma]=\int\mathcal{D}^{NS}\xi\;\;\tilde{F}_{i_{\pi}}(\xi_{\pi})\;
  {\prod\limits_{p}}^{\frac{NS}{2}}
  F_{i_p}(\xi_{-p},\xi_p)\;{\prod\limits_{j=1}^L}\;e^{\,\sigma_j\xi_j},
  \eb
  where all the indices $\{i_p\}$, except $i_{\pi}$, take on
  four values as above, and the functions $F_1\ldots F_4$ are
  given by (\ref{efs})--(\ref{endefs}). The index $i_{\pi}$ can
  have only two values, 1 and 2, and the corresponding functions
  $\tilde{F}_1$ and $\tilde{F}_2$ are simply
  \ben
  \tilde{F}_1(\xi)=1,\qquad\qquad   \tilde{F}_2(\xi)=\xi.
  \ebn
  One may verify that the function (\ref{ansatz2}) gives an
  eigenvector of $T^{NS}$ with the eigenvalue
  \ben
  \Lambda^{NS}_{\{i_p\}}=2^L\,\zeta^{NS}\,\tilde{\lambda}_{i_{\pi}}(\pi)
  {\prod\limits_{p}}^{\frac{NS}{2}}\lambda_{i_p}(p),
  \ebn
  where all $\lambda_{i}(p)$ are defined as above and
  \ben
  \tilde{\lambda}_1(\pi)=1,\qquad\tilde{\lambda}_2(\pi)=G_{12}(\pi)=\frac{v_{\pi}}{4\chi_{\pi}}\,.
  \ebn
  Since the total number of eigenvectors (\ref{ansatz2}) is equal
  to $2\times 4^{(L-1)/2}=2^L$, the diagonalization of $T^{NS}$ is
  completed.

  The first thing that may seem unusual is that if we set $i_p=1$
  for all $p$, including $p=\pi$, the corresponding eigenvector of
  $T^{NS}$ will not always represent the physical vacuum.
  Moreover, this vector is \textit{odd}
  under spin reflection and, therefore, it is not an
  eigenvector of the full transfer matrix~$T$. Note also that
  \ben
  \rho_{\pi}^{1/2}=2\chi_{\pi}^{1/2}\max\left\{\frac{|v_{\pi}|}{4\chi_{\pi}},1\right\}.
  \ebn
  Therefore, if one tries to write the eigenvalues in the form,
  analogous to (\ref{eigenphys1}), then the result will be different in
  different regions of parameters. Namely, for
  $|v_{\pi}|/4\chi_{\pi}\geq1$ one obtains
  \ben
  \Lambda_{\{i_p\}}^{NS}=\Lambda^{NS}_{max}\,\left(\frac{v_{\pi}}{\rho_{\pi}}\right)^{2-i_{\pi}}\!\!\!
  {\prod\limits_{\;\;p|i_p=2}}^{\!\!\!\frac{NS}{2}}\;\frac{v_p}{\rho_p}
  \;{\prod\limits_{\;\;p|i_p=3}}^{\!\!\!\frac{NS}{2}}\;\frac{v_{-p}}{\rho_p}
  \;{\prod\limits_{\;\;p|i_p=4}}^{\!\!\!\frac{NS}{2}}\;\frac{v_p\,v_{-p}}{\rho^{\,2}_p}\,,
  \ebn
  and for $|v_{\pi}|/4\chi_{\pi}\leq1$ we have
  \ben
  \Lambda_{\{i_p\}}^{NS}=\Lambda^{NS}_{max}\,\left(\frac{v_{\pi}}{\rho_{\pi}}\right)^{i_{\pi}-1}\!\!\!
  {\prod\limits_{\;\;p|i_p=2}}^{\!\!\!\frac{NS}{2}}\;\frac{v_p}{\rho_p}
  \;{\prod\limits_{\;\;p|i_p=3}}^{\!\!\!\frac{NS}{2}}\;\frac{v_{-p}}{\rho_p}
  \;{\prod\limits_{\;\;p|i_p=4}}^{\!\!\!\frac{NS}{2}}\;\frac{v_p\,v_{-p}}{\rho^{\,2}_p}\,,
  \ebn
  where $\Lambda^{NS}_{max}$ is defined by the formula
  (\ref{lnsm}). Thus one can again interpret the eigenvectors of
  $T^{NS}$ as multiparticle states
  $|p_1,\ldots,p_k\rangle_{NS}$. The main differences with the previous
  case are the following:
  \begin{itemize}
  \item If $|v_{\pi}|/4\chi_{\pi}\geq1$, then the states, containing a
  particle with the momentum $p=\pi$, are given by the ansatz
  (\ref{ansatz2}) with $i_{\pi}=1$; for
  $|v_{\pi}|/4\chi_{\pi}\leq1$ they correspond to the choice
  $i_{\pi}=2$.
  \item For $|v_{\pi}|/4\chi_{\pi}\geq1$ the eigenstates, which
  are even (odd) under spin reflection, contain even (odd) number
  of particles, while for $|v_{\pi}|/4\chi_{\pi}\leq1$ this number
  should be odd (even).
  \end{itemize}
  \subsection{R-sector, odd $L$}
  The treatment of this case is completely analogous to the
  previous one, since for odd $L$  Ramond spectrum contains only one
  `special' mode $p=0$. All the eigenvectors and eigenvalues of the matrix $T^R$
  are given by
  \be\label{ansatz3}
  f^{\,R}_{\{i_p\}}[\sigma]=\int\mathcal{D}^R\xi\;\;\tilde{F}_{i_{0}}(\xi_{0})\;
  {\prod\limits_{p}}^{\frac{R}{2}}
  F_{i_p}(\xi_{-p},\xi_p)\;{\prod\limits_{j=1}^L}\;e^{\,\sigma_j\xi_j},
  \eb
  \ben
  \Lambda^{\,R}_{\{i_p\}}=2^L\,\zeta^{R}\,\tilde{\lambda}_{i_{0}}(0)
  {\prod\limits_{p}}^{\frac{R}{2}}\lambda_{i_p}(p).
  \ebn
  Here the indices $\{i_p\}_{p\neq0}$ take on four values,
  $i_0=1,2$, the functions $\{\tilde{F}_j\}$,
  $\{F_j\}$, $\{\lambda_j\}$ are
  defined as above and
  \ben
  \tilde{\lambda}_1(0)=1,\qquad \tilde{\lambda}_2(0)=G_{12}(0)=\frac{v_{0}}{4\chi_{0}}\,.
  \ebn

  Again, since we have
  \ben
  \rho_{0}^{1/2}=2\chi_{0}^{1/2}\max\left\{\frac{|v_{0}|}{4\chi_{0}},1\right\},
  \ebn
  the quasiparticle interpretation of eigenvalues and eigenvectors
  is different in the regions $|v_{0}|/4\chi_{0}\geq1$ and
  $|v_{0}|/4\chi_{0}\leq1$. Namely, one has
  \ben
  \Lambda_{\{i_p\}}^{R}=\Lambda^{R}_{max}\,\left(\frac{v_{0}}{\rho_{0}}\right)^{2-i_{0}}\!\!\!
  {\prod\limits_{\;\;p|i_p=2}}^{\!\!\!\frac{R}{2}}\;\frac{v_p}{\rho_p}
  \;{\prod\limits_{\;\;p|i_p=3}}^{\!\!\!\frac{R}{2}}\;\frac{v_{-p}}{\rho_p}
  \;{\prod\limits_{\;\;p|i_p=4}}^{\!\!\!\frac{R}{2}}\;\frac{v_p\,v_{-p}}{\rho^{\,2}_p}\,
  \quad \text{for}\;\;|v_{0}|/4\chi_{0}\geq1,
  \ebn
  \ben
  \Lambda_{\{i_p\}}^{R}=\Lambda^{R}_{max}\,\left(\frac{v_{0}}{\rho_{0}}\right)^{i_{0}-1}\!\!\!
  {\prod\limits_{\;\;p|i_p=2}}^{\!\!\!\frac{R}{2}}\;\frac{v_p}{\rho_p}
  \;{\prod\limits_{\;\;p|i_p=3}}^{\!\!\!\frac{R}{2}}\;\frac{v_{-p}}{\rho_p}
  \;{\prod\limits_{\;\;p|i_p=4}}^{\!\!\!\frac{R}{2}}\;\frac{v_p\,v_{-p}}{\rho^{\,2}_p}\,
  \quad \text{for}\;\;|v_{0}|/4\chi_{0}\leq1,
  \ebn
 where the eigenvalue with the maximum modulus,
 $\Lambda^{\,R}_{max}$, is given by
 \ben
 \Lambda^{\,R}_{max}={\prod\limits_p}^R\rho^{1/2}_{p}.
 \ebn
  Similarly to the above, let us denote by
  $|p_1,\ldots,p_k\rangle_R$ the eigenstate of $T^R$, containing
  $k$  particles with distinct R-momenta $p_1,\ldots,p_k$.

  Note that  even (odd) eigenstates of $T^R$ should contain
  even (odd) number of particles for $|v_{0}|/4\chi_{0}\geq1$, and odd (even) number
  of particles for $|v_{0}|/4\chi_{0}\leq1$. This change can be
  easily understood if we take, say, $|v_0|/4\chi_0\leq1$, and
  consider two eigenstates,  $|p_1,\ldots,p_k\rangle_R$ and  $|0,p_1,\ldots,p_k\rangle_R$
  ($p_j\neq0$, $j=1,\ldots,k$). Then let us gradually increase
  $|v_0|/4\chi_0$. When this parameter approaches the critical
  value~1, two eigenstates correspond to the same eigenvalue, and
  when it exceeds 1, the roles of two vectors swap around: the
  particle with zero momentum disappears from the second vector
  (thus decreasing the number of particles by 1) and appears in
  the first (the number of particles increases by 1).

  \subsection{R-sector, even $L$}
  Since for even $L$ the Ramond spectrum of quasimomenta contains
  both $p=0$ and $p=\pi$, the eigen\-vectors and
  eigenvalues of $T^R$ in this case can be written in the following way:
  \be\label{ansatz4}
  f^{\,R}_{\{i_p\}}[\sigma]=\int\mathcal{D}^R\xi\;\;
  \tilde{F}_{i_{0}}(\xi_{0})\tilde{F}_{i_{\pi}}(\xi_{\pi})\;
  {\prod\limits_{p}}^{\frac{R}{2}}
  F_{i_p}(\xi_{-p},\xi_p)\;{\prod\limits_{j=1}^L}\;e^{\,\sigma_j\xi_j},
  \eb
  \ben
  \Lambda^{\,R}_{\{i_p\}}=2^L\,\zeta^{R}\,\tilde{\lambda}_{i_{0}}(0)\tilde{\lambda}_{i_{\pi}}(\pi)
  {\prod\limits_{p}}^{\frac{R}{2}}\lambda_{i_p}(p)\,.
  \ebn
  From the physical point of view, here one should distinguish four
  different regions in the space of parameters. They have the
  following properties:
  \begin{itemize}
  \item
  ${\underline{|v_0|\geq4\chi_0,\;|v_{\pi}|\geq4\chi_{\pi}}}$. The
  eigenstates of $T^R$, containing a particle with the momentum
  $p=0$ ($p=\pi$), are given by the formula (\ref{ansatz4}) with
  $i_0=1$ ($i_{\pi}=1$). The eigenvectors, which are even (odd) under
  spin reflection, should
  contain even (odd) number of particles.
   \item
  ${\underline{|v_0|\geq4\chi_0,\;|v_{\pi}|\leq4\chi_{\pi}}}$. The
  eigenstates, containing a particle with the momentum
  $p=0$ ($p=\pi$), correspond to
  $i_0=1$ ($i_{\pi}=2$). Even (odd) eigenvectors
  contain odd (even) number of particles.
  \item
  ${\underline{|v_0|\leq4\chi_0,\;|v_{\pi}|\geq4\chi_{\pi}}}$. Particle
   with the momentum  $p=0$ ($p=\pi$) corresponds to
  $i_0=2$ ($i_{\pi}=1$). Even (odd) eigenvectors
  contain odd (even) number of particles.
 \item
  ${\underline{|v_0|\leq4\chi_0,\;|v_{\pi}|\leq4\chi_{\pi}}}$.
  Particle
   with the momentum  $p=0$ ($p=\pi$) corresponds to
  $i_0=2$ ($i_{\pi}=2$). Even (odd) eigenvectors
  contain even (odd) number of particles.
  \end{itemize}

  \section{Norms and form factors}
  In the present section, the problem of computation of
  correlation functions of the BBS$_2$ model is addressed.
  Local fields will be represented by spin variables $\sigma_{i,j}$
  ($i=1,\ldots,L$; $j=1,\ldots,M$). In the transfer matrix
  formalism, $2k$-point correlation functions
  $\langle\sigma_{i_1,j_1}\sigma_{i_2,j_2}\ldots\sigma_{i_{2k},j_{2k}}\rangle$
  can be written in the following way\footnote{All $(2k+1)$-point correlation
  functions vanish due to $\mathbb{Z}_2$-symmetry of the model.}:
  \be\label{cf000}
  \;\langle\sigma_{i_1,j_1}\ldots\sigma_{i_{2k},j_{2k}}\rangle=\eb
  \ben
  =Z^{-1}(L,M)\sum\limits_{[\sigma^{(1)}]}\ldots\sum\limits_{[\sigma^{(2k)}]}
  \sigma^{(1)}_{i_1}\;T^{j_2-j_1}[\sigma^{(1)},\sigma^{(2)}]\;
  \sigma^{(2)}_{i_2}\;T^{j_2-j_1}[\sigma^{(2)},\sigma^{(3)}]\ldots
  \sigma^{(2k)}_{i_{2k}}\;T^{M-(j_{2k}-j_1)}[\sigma^{(2k)},\sigma^{(1)}]\,,
  \ebn
  where it was assumed that $j_1\leq j_2\leq\ldots j_{2k}$.  Let us introduce spin operator
  \ben
  S_{1,1}[\sigma,\sigma']=\sigma_{1}\;\delta_{[\sigma],\,[\sigma']}=
  \sigma_1\,\prod\limits_{j=1}^L\frac{1+\sigma_j\,\sigma'{}_{\!\!j}}{2},
  \ebn
  acting on functions $f[\sigma]\in V$ from the left in the usual way. If we
  make use of the translation operator $R$ (see formula
  (\ref{translation})) to define
  \ben
  \qquad\qquad S_{i,j}=T^{j-1}\,R^{i-1}\,S_{1,1}\,R^{1-i}\,T^{1-j},\qquad
  i=1,\ldots,L,\quad j=1,\ldots,M,
  \ebn
  then one may rewrite (\ref{cf000}) as
  \be\label{cf00}
  \langle\sigma_{i_1,j_1}\ldots\sigma_{i_{2k},j_{2k}}\rangle=
  \frac{\mathrm{Tr}\left(S_{i_1,j_1}S_{i_2,j_2}\ldots
  S_{i_{2k},j_{2k}}\,T^M\right)}{\mathrm{Tr}\,T^M}\,.
  \eb
  Since all the eigenvalues of the transfer matrix $T$ are known, the problem reduces
  to the calculation of the trace in the numerator. One would want
  to compute this trace in the basis of eigenstates of $T$.
  However, such computation is not quite straightforward, since the transfer
  matrix of the BBS$_2$ model is not symmetric and thus its eigenvectors
  are not necessarily orthogonal. Therefore, in order to construct the dual basis,
  one should separately find the eigenvectors for the \textit{right} action of $T$,
  (since $T$ is
  not symmetric, they can not be obtained from the eigenvectors,
  found in the previous section, by simple transposition).

  Assume for a moment that we have found all
  `left' and `right' eigenvectors of $T$. Let us denote them by
  $|n\rangle$ and $\langle n|$ , where $n$ is any convenient set of
  quantum numbers, identifying the eigenstate (for example, the
  number of particles  and their quasimomenta). The resolution of
  the identity matrix in this basis of eigenstates has the form
  \be\label{resolution}
  \mathbf{1}=\sum\limits_{n}\, b_{n}\,|n\rangle\langle n|, \qquad
  b_n=1/\langle n|n\rangle.
  \eb
  The relation (\ref{resolution}) means, in particular, that the
  trace of any matrix $X$ can be written as
  \ben
  \mathrm{Tr}\,X=\sum\limits_{n}b_{n}\langle n|X|n\rangle\,.
  \ebn

  Recall also that the eigenvectors of $T$ diagonalize as well the
  translation operator $R$. Therefore, inserting the
  resolution of the identity matrix into the representation
  (\ref{cf00}) $k$ times, one can rewrite $2k$-point correlation
  function in the form of the so-called form factor expansion. For
  example, for the 2-point correlation function one has
  \be\label{2pointff}
  \langle\sigma_{i_1,j_1}\sigma_{i_2,j_2}\rangle=
  \frac{\sum\limits_{m,n}\;b_{m}\,b_{n}\,
  \langle n|S_{1,1}|m\rangle\langle m|S_{1,1}|n\rangle\,
  e^{-E_m(j_2-j_1)-E_n(M-j_2+j_1)+i(P_m-P_n)(i_2-i_1)}}{\sum\limits_n
  e^{-ME_n}}\,.
  \eb
  Matrix elements $\langle n|S_{1,1}|m\rangle$, entering this
  formula, hereinafter will be referred to as form factors. Parameters $E_n$ and $P_n$
  have the meaning of energy and total momentum of the state $|n\rangle$
  (and $\langle n|$). They are related to the eigenvalues of $T$
  and $R$ in the following way:
  \ben
  T|n\rangle=\Lambda_{max}\,e^{-E_n}|n\rangle,\qquad
  R|n\rangle=e^{iP_n}|n\rangle,
  \ebn
  where $\Lambda_{max}$ denotes the eigenvalue of $T$ with the
  maximum modulus. In the BBS$_2$ model, the values of $E_n$ and
  $P_n$, corresponding to the multiparticle state
  $|n\rangle=|p_1,\ldots,p_k\rangle$, are given by the sums of
  one-particle energies and momenta.

  The generalization of the form factor expansion (\ref{2pointff})
  to the multipoint case is straightforward. Thus in order to find
  all correlation functions, only three further steps should be
  made.
  First one should find the eigenvectors for the right
  action of the transfer matrix $T$, i.~e. the functions $f[\sigma]\in
  V$ such that $\sum\limits_{[\sigma]}f[\sigma]T[\sigma,\sigma']=\lambda_f
  f[\sigma']$.
  Then one needs to compute  scalar products  $\langle n|n\rangle$.
  Finally, the most difficult task is the calculation of form factors $\langle n|S_{1,1}|m\rangle$.
  All these problems are treated
  (the third one with only a partial success) in the following subsections.
  \subsection{Eigenvectors for the right action of $T$}
  The variables $[\sigma]$ and $[\sigma']$ enter into the
  representation (\ref{tnsrf}) for the matrices $T^{NS}$ and $T^R$
  in a similar way. Therefore, one may construct the eigenvectors for
  the right action of $T$ along the lines of Section~3. However,
  there exists even more straightforward way to obtain them.
  Note that the right action of $T$ on $f[\sigma]\in
  V$ coincides with the left action of the transfer matrix
  $\dot{T}=T^T$ of another BBS$_2$ model (see Fig.~1b), characterized by the
  parameters
  \be\label{newparams}
  \dot{a}_{12}=a_{12},\qquad
  \dot{a}_{13}=a_{24},\qquad\dot{a}_{14}=a_{23},\qquad
  \dot{a}_{23}=a_{14},\qquad\dot{a}_{24}=a_{23},\qquad\dot{a}_{34}=a_{34}.
  \eb
  Thus the eigenvectors we are looking for may be obtained
  from already found ones by the substitution (\ref{newparams})
  and mathching the eigenvalues.

  It is easy to verify that under
  the above substitution various quantities, used in the
  construction of eigenvectors and eigenvalues, change as follows:
  \ben
  \left(\begin{array}{cc}
  G_{11}(p) & G_{12}(p) \\
  G_{21}(p) & G_{22}(p) \end{array}\right)\rightarrow
   \left(\begin{array}{cc}
  \dot{G}_{11}(p) & \dot{G}_{12}(p) \\
  \dot{G}_{21}(p) & \dot{G}_{22}(p) \end{array}\right)=
   \left(\begin{array}{cc}
  G_{22}(p) & G_{21}(p) \\
  G_{12}(p) & G_{11}(p) \end{array}\right),
  \ebn
  \be\label{apmpoint}
  A^{\pm}(p)\rightarrow  \dot{A}^{\pm}(p)=
  \frac{1+G(p)
  \mp\sqrt{\bigl(1-G(p)\bigr)^2-4G_{12}(p)G_{21}(p)}}{2\,G_{11}(p)}\,,
  \eb
  \ben
  \chi_p\rightarrow\dot{\chi}_p=\chi_p\,,\qquad u_p\rightarrow\dot{u}_p=u_p\,,\qquad
  v_p\rightarrow\dot{v}_p=v_{-p}\,.
  \ebn
  Let us now consider, for instance, NS-sector and assume that $L$
  is even. Let $f_{\{i_p\}}^{NS}[\sigma]$ denote the `left'
  eigenvector (\ref{ansatz1}), corresponding to the eigenvalue
  $\Lambda^{NS}_{\{i_p\}}$. Under the substitution
  (\ref{newparams}) `partial' eigenvalues $\lambda_2(p)$ and
  $\lambda_3(p)$ (formulas (\ref{lambdas02}), (\ref{lambdas03}))
  exchange their roles, while $\lambda_1(p)$ and $\lambda_4(p)$
  remain unchanged. Then it becomes clear that the `right'
  eigenvector $\dot{f}^{NS}_{\{i_p\}}[\sigma]$ of $T^{NS}$,
  corresponding to the same eigenvalue as
  $f_{\{i_p\}}^{NS}[\sigma]$, is given by
  \be\label{ansatz1r}
  \dot{f}^{NS}_{\{i_p\}}[\sigma]=\int\mathcal{D}^{NS}\dot{\xi}\;\;{\prod\limits_{p}}^{\frac{NS}{2}}
  \dot{F}_{i_p}(\dot{\xi}_{-p},\dot{\xi}_p)\;{\prod\limits_{j=1}^L}\;e^{\,\sigma_j\dot{\xi}_j},
  \eb
  with
  \begin{eqnarray}\label{efsr1}
  \dot{F}_1(\dot{\xi}_{-p},\dot{\xi}_p)&=&\exp\Bigl(\dot{\xi}_{-p}\,\dot{A}^+(p)\,\dot{\xi}_p\Bigr),\\
  \label{efsr2} \dot{F}_2(\dot{\xi}_{-p},\dot{\xi}_p)&=&\dot{\xi}_{p}\,,\\
  \label{efsr3}\dot{F}_3(\dot{\xi}_{-p},\dot{\xi}_p)&=&\dot{\xi}_{-p}\,,\\
  \label{efsr4}\dot{F}_4(\dot{\xi}_{-p},\dot{\xi}_p)&=&
  \exp\Bigl(\dot{\xi}_{-p}\,\dot{A}^-(p)\,\dot{\xi}_p\Bigr),
  \end{eqnarray}
  the functions $\dot{A}^{\pm}(p)$ being defined by the formula
  (\ref{apmpoint}).
  Dotted grassmann variables $\dot{\xi}$ are used in the
  representation (\ref{ansatz1r}) for further convenience in the
  computation of scalar products and form factors.

  In order to obtain a similar answer for the other cases (Neveu-Schwartz
  sector for odd $L$ and Ramond sector), it is sufficient to
  substitute  in (\ref{ansatz2}), (\ref{ansatz3}) and (\ref{ansatz4}) instead
  of $F_1\ldots F_4$ new functions $\dot{F}_1\ldots\dot{F}_4$.
  The functions $\tilde{F}_1$ and $\tilde{F}_2$, which are
  responsible for the special modes $p=0,\pi$, remain unchanged.

  \subsection{Normalization}
  It is instructive to consider not only the norms $\langle
  n|n\rangle$, but also general scalar products $\langle
  m|n\rangle$, and to verify by hand that $\langle
  m|n\rangle=0$ for $m\neq n$. First one should remark that the
  eigenvectors of $T$, which belong to different sectors, are
  orthogonal, since they correspond to different eigenvalues of
  $U$. Thus one may look at each sector separately. Let us now
  consider, for instance, the Neveu-Schwartz sector for even $L$.
  Let us take a `right' eigenvector
  $\bigr._{NS}\bigl\langle\{i_p\}\bigr|\;\substack{\text{\textit{def}}\\=\\ \;}\;
  \dot{f}^{NS}_{\{i_p\}}[\sigma]$
  (given by the formula (\ref{ansatz1r}))
  and a `left' eigenvector
  $\bigl|\{j_p\}\bigr\rangle_{NS}\;\substack{\text{\textit{def}}\\=\\ \;}\;f^{NS}_{\{j_p\}}[\sigma]$
   (given by the formula (\ref{ansatz1})), and then compute their
   scalar product
   \ben
   \bigr._{NS}\bigl\langle\{i_p\}\bigr|\{j_p\}\bigr\rangle_{NS}=\sum\limits_{[\sigma]}
   \,\dot{f}^{NS}_{\{i_p\}}[\sigma]\;f^{NS}_{\{j_p\}}[\sigma].
   \ebn
   Since the fields $\xi$ and $\dot{\xi}$ in the representations
   (\ref{ansatz1}) and (\ref{ansatz1r}) commute, the summation over
   intermediate spins can be easily done and one obtains
   \ben
   \bigr._{NS}\bigl\langle\{i_p\}\bigr|\{j_p\}\bigr\rangle_{NS}=
   2^L\int\mathcal{D}^{NS}\xi\,\mathcal{D}^{NS}\dot{\xi}\;\;
   {\prod\limits_{p}}^{\frac{NS}{2}}\left(\,
  \dot{F}_{i_p}(\dot{\xi}_{-p},\dot{\xi}_p)\;F_{j_p}(\xi_{-p},\xi_p)\;
  e^{\,\dot{\xi}_{-p}\,\xi_p\,+\,\xi_{-p}\,\dot{\xi}_{p}}\right).
   \ebn
   Calculation of this factorized integral gives
   \ben
   \bigr._{NS}\bigl\langle\{i_p\}\bigr|\{j_p\}\bigr\rangle_{NS}
   =2^L\;{\prod\limits_{p}}^{\frac{NS}{2}}\alpha_{i_p j_p}(p),
   \ebn
   where the functions $\alpha_{ij}(p)$ can be assembled into a
   $4\times4$ matrix
   \ben
   \bigl\|\alpha_{ij}(p)\bigr\|_{i,j=1,\ldots,4}=
   \left(\begin{array}{crrc}
   \dot{A}^+(p)A^+(p)-1 & 0 & 0 & \dot{A}^+(p)A^-(p)-1 \\
   0 & -1 & 0 & 0 \\
   0 & 0 & -1 & 0 \\
   \dot{A}^-(p)A^+(p)-1 & 0 & 0 &  \dot{A}^-(p)A^-(p)-1
   \end{array}\right).
   \ebn
   Using the explicit formulas for $\dot{A}^{\pm}(p)$ and
   $A^{\pm}(p)$, one may check that $
   \dot{A}^{\pm}(p)A^{\mp}(p)=1$. Therefore, `right' and `left'
   eigenvectors, corresponding to different eigenvalues, are
   orthogonal (as it should be). The norm
   $\bigr._{NS}\bigl\langle\{i_p\}\bigr|\{i_p\}\bigr\rangle_{NS}$
   is given by
   \be\label{norm}
   \bigr._{NS}\bigl\langle\{i_p\}\bigr|\{i_p\}\bigr\rangle_{NS}=
   2^L\;{\prod\limits_{p}}^{\frac{NS}{2}}\alpha_{i_p}(p),
   \eb
   where we have introduced the notation
   \ben
   \alpha_1(p)=1- \dot{A}^+(p)A^+(p),\qquad
   \alpha_2(p)=\alpha_3(p)=1,\qquad \alpha_4(p)=1-
   \dot{A}^-(p)A^-(p),
   \ebn
   and corrected the overall sign. One may check that the answer for
   the Neveu-Schwartz sector and odd $L$ is given by the same formula (\ref{norm}).
   The only things that change in the Ramond sector are the values of
   quasimomenta.
  \subsection{Form factors}
  Since the eigenstates of $T$ from the same sector are all simultaneously even or odd under
  spin reflection,  all form factors of type NS--NS and R--R are equal to zero.
  Now assume for definiteness that $L$ is even and consider
  a `right' eigenvector
  $\bigr._{NS}\bigl\langle\{i_p\}\bigr|=\dot{f}^{NS}_{\{i_p\}}[\sigma]$
  from the Neveu-Schwartz sector and a `left' eigenvector
  $\bigr|\{j_p\}\bigr\rangle_{R}={f}^{R}_{\{j_p\}}[\sigma]$ from
  the Ramond sector. Let us calculate the form factor
  \ben
  \bigr._{NS}\bigl\langle\{i_p\}\bigr|S_{1,1}\bigr|\{j_p\}\bigr\rangle_{R}=
  \sum\limits_{[\sigma]}\,\sigma_1\,\dot{f}^{NS}_{\{i_p\}}[\sigma]\,{f}^{R}_{\{j_p\}}[\sigma]\,.
  \ebn
  After summation over intermediate spins one obtains
  \be\label{ffint}
  \bigr._{NS}\bigl\langle\{i_p\}\bigr|S_{1,1}\bigr|\{j_p\}\bigr\rangle_{R}=
  \eb
  \ben
  =2^L\int\mathcal{D}^R\xi\,\mathcal{D}^{NS}\dot{\xi}\;\;
  \tilde{F}_{j_0}(\xi_0)\tilde{F}_{j_{\pi}}(\xi_{\pi})\;
  {\prod\limits_{p}}^{\frac{R}{2}}
  {F}_{j_p}({\xi}_{-p},{\xi}_p)\;
  {\prod\limits_{q}}^{\frac{NS}{2}}
  \dot{F}_{i_q}(\dot{\xi}_{-q},\dot{\xi}_q)\;\bigl(\xi_1+\dot{\xi}_1\bigr)\;
  \exp\left\{\sum\limits_{k=1}^{L}\xi_k\,\dot{\xi}_k\right\}.
  \ebn

  Unfortunately, we have not managed to find a compact expression
  for this gaussian integral, although we strongly suspect it is
  possible. In order to illustrate emerging difficulties, let us
  assume that $|v_0|<4\chi_0$, $|v_{\pi}|>4\chi_{\pi}$ (this region mimics
  ferromagnetic phase),
  and consider the simplest possible form factor
  $_{NS}\langle vac|S_{1,1}|vac\rangle_{R}$, which corresponds to the
  following choice: $j_0=1$,
  $j_{\pi}=2$, $j_p=1$ for all $p\in(0,\pi)$, $i_q=1$ for all
  $q\in(0,\pi)$. One then obtains
  \ben
  _{NS}\langle vac|S_{1,1}|vac\rangle_{R}=\ebn\ben=2^L\!\!
  \int\!\mathcal{D}^{R}\xi\,\mathcal{D}^{NS}\dot{\xi}\,\;\xi_{\pi}\,(\xi_1+\dot{\xi}_1)
  \exp\left\{{\sum\limits_p}^{\frac{R}{2}}\xi_{-p\;}A^+(p)\xi_p+
  {\sum\limits_q}^{\frac{NS}{2}}\dot{\xi}_{-q\;}\dot{A}^+(q)\dot{\xi}_q+
  \sum\limits_{k=1}^{L}\xi_k\,\dot{\xi}_k\right\}.
  \ebn
  Quadratic form in the exponential consists of three pieces,
  which can not be diagonalized simultaneously: the first and the
  second piece are diagonal in the Fourier basis with Ramond and
  Neveu-Schwartz values of discrete quasimomenta, and the third
  one is diagonal in the coordinate representation. Actually, one
  can now remove the dots, using the following rule: all quadratic
  terms in the exponential, containing a dotted variable on the
  left, should change their signs. Performing this operation and
  passing to the coordinate representation in all terms, one
  obtains
  \be\label{ffi}
  _{NS}\langle vac|S_{1,1}|vac\rangle_{R}=\frac{2^L}{\sqrt{L}}
  \int\!\mathcal{D}\xi\,\mathcal{D}\eta\,\;\sum\limits_{k=1}^L
  (-1)^k\xi_k\,(\xi_1+\eta_1)\;\exp\left\{
  \frac12\left(\begin{array}{rr}\xi & \eta\end{array}\right)
  \left(\begin{array}{cc}
  A^+ & \mathbf{1} \\ \!\!\!\!-\mathbf{1} & -\dot{A}^+\end{array}\right)
  \left(\begin{array}{c} \xi \\ \eta \end{array}\right)\right\},
  \eb
  where antisymmetric $L\times L$ matrices $A^+$, $\dot{A}^+$ are
  given by
  \ben
  A^{+}_{xx'}=\frac{1}{L}{\sum\limits_{\;\;p\neq 0,\pi}}^{\!\!\!R}A^{+}(p)\,e^{ip(x-x')},\qquad
  \dot{A}^{+}_{xx'}=\frac{1}{L}\;{\sum\limits_{q}}^{NS}\dot{A}^{+}(q)\,e^{iq(x-x')},
  \qquad x,x'=1,\ldots,L.
  \ebn
  Evaluation of the gaussian integral (\ref{ffi}) gives
  \be\label{ffh}
  _{NS}\langle vac|S_{1,1}|vac\rangle_{R}=\frac{2^L}{\sqrt{L}}\;
  \mathrm{Pf}(\dot{A}^+)\,\mathrm{Pf}(H)\,
  \sum\limits_{k=1}^L
  \,(-1)^k
  \left[H^{-1}-\left(\dot{A}^+\right)^{-1}H^{-1}\right]_{1k},
  \eb
  where $L\times L$ matrix $H$ is also antisymmetric and has
  Toeplitz form:
  \ben
  H_{xx'}=\left(\dot{A}^+\right)^{-1}_{xx'}-A^{+}_{xx'}=
  \frac{1}{L}\;{\sum\limits_{q}}^{NS}{A}^{-}(q)\,e^{iq(x-x')}-
  \frac{1}{L}{\sum\limits_{\;\;p\neq
  0,\pi}}^{\!\!\!R}A^{+}(p)\,e^{ip(x-x')}.
  \ebn
  Thus the problem of computation of the form factor
  $_{NS}\langle vac|S_{1,1}|vac\rangle_{R}$ is reduced to the
  calculation of the determinant of $H$ and inverse matrix
  $H^{-1}$. Actually, this is also the case for more complicated
  form factors. In spite of the remarkably simple form of the
  matrix $H$, we have not succeded in the calculation of $\mathrm{Pf}(H)$
  and $H^{-1}$. However, we believe that the representations of
  type (\ref{ffh}) are still useful, since they effectively reduce
  initial $2^L$-dimensional problem to an $L$-dimensional one.

  It should also be pointed out that form factor $_{NS}\langle vac|S_{1,1}|vac\rangle_{R}$
  enters into the definition of the order
  parameter of the BBS$_2$ model. More precisely, one has
  \ben
  \langle\sigma\rangle^2\;\;\substack{\text{\textit{def}}\\=\\
  \;}\;\lim_{i,j\rightarrow\infty}\left(\lim_{L,M\rightarrow\infty}\langle\sigma_{1,1}\sigma_{i,j}\rangle
  \right)=
  \lim_{L\rightarrow\infty}\frac{_{NS}\langle vac|S_{1,1}|vac\rangle_{R}
  \,_{R}\langle vac|S_{1,1}|vac\rangle_{NS}}{_{NS}\langle
  vac|vac\rangle_{NS}\;
  _{R}\langle vac|vac\rangle_{R}\;\;\;\;}\,.
  \ebn
  Although we have not managed to obtain a closed expression for
  this form factor, the order parameter can presumably be
  calculated by another method. We hope to return to this problem elsewhere.

  \section{Special cases}
  \subsection{BBS$_2$ model}
  Parameters of the general free-fermion model, which correspond
  to BBS$_2$ model (via the formulas (\ref{bbspars})--(\ref{bbsparsf})),
  are not independent. In particular, in addition to free-fermion
  condition (\ref{ffc}), they also satisfy the relation
  $a_{13}a_{24}=a_{14}a_{23}$. Therefore, one could expect some
  simplifications of the above formulas for tranfer matrix eigenvectors to occur in this case.
  Furthermore, it is known that the eigenvalues of the BBS$_2$ transfer matrix
  should have polynomial dependence on spectral variable $t$,
  and that the eigenvectors should not depend on it. In order to verify
  these properties, let us rewrite our formulas in the BBS
  notation.

  The variables $\chi_p$ and $G_{ij}(p)$ ($i,j=1,2$), which were
  used in the grassmann integral representation of the transfer
  matrix, are expressed in terms of $t,x,x',y,y',\mu,\mu'$ as
  \ben
  \chi_p=\frac{4(t+\mu\mu' xx')^2+4(yy'-t\mu\mu')^2+8(t+\mu\mu'xx')(yy'-t\mu\mu')\cos p}
  {(y+\mu t)(y'+\mu')}\,,
  \ebn
  \ben
  \chi_p\, G_{11}(p)=\frac{8it(y+\mu\mu'x')(y'-\mu\mu'x)\sin p}{(y+\mu
  t)^2(y'+\mu')^2}\,\ebn
  \ben
    \chi_p\, G_{22}(p)=\frac{8it(y-\mu\mu'x')(y'+\mu\mu'x)\sin p}{(y+\mu
  t)^2(y'+\mu')^2}\,,
  \ebn
  \ben
  \chi_p\, G_{12}(p)=
  \frac{16(t^2+\mu^2\mu'^2(t^2-x^2x'^2)-y^2y'^2)-32\mu\mu'(t^2+xx'yy')\cos
  p -32it\mu\mu'(xy+x'y')\sin p}{(y+\mu
  t)^2(y'+\mu')^2}\,.
  \ebn
  In order to write down the eigenvalues of $T^{NS}$ and $T^{R}$,
  it is sufficient to express in terms of BBS parameters the
  quantities $\Lambda^{NS(R)}_{max}$ and ${v_p}/{\rho_p}\,$.
  They are given by (see also \cite{bis})
  \ben
  \Lambda^{NS}_{max}=\frac{(1+\mu^L\mu'^L)}{y^L
  y'^L}\;{\prod\limits_{p}}^{NS}(t+t_p),\ebn\ben
 \Lambda^{R}_{max}=\frac{(1-\mu^L\mu'^L)}{y^L
  y'^L}\;{\prod\limits_{p}}^{R}(t+t_p),
  \ebn\ben
  {v_p}/{\rho_p}=\frac{t-t_p}{t+t_p},
  \ebn
  where $t_p$ is defined as
  \ben
  t_p=\frac{\sqrt{a_p c_p-b_p^2}-ib_p}{a_p},
  \ebn
  with
  \ben
  a_p=1-2\mu\mu'\cos p+\mu^2\mu'^2,\ebn
  \ben
  b_p=\mu\mu'(xy+x'y')\sin p.
  \ebn
  \ben
  c_p=y^2y'^2+2\mu\mu'xx'yy'\cos p+\mu^2\mu'^2x^2x'^2.\ebn
  Finally, `right' and `left' eigenvectors of $T^{NS}$ and $T^R$ are fully
  characterized by the functions $A^{\pm}(p)$ and $\dot{A}^{\pm}(p)$,
  which in the case of the BBS$_2$ model can be written in the
  following form:
  \begin{eqnarray*}
  A^{\pm}(p)&=&\frac{d_p\mp\sqrt{a_p c_p-b_p^2}}{(y-\mu\mu'x')(y'+\mu\mu'x)\,i\sin
  p}\,,\\
  \dot{A}^{\pm}(p)&=&\frac{d_p\mp\sqrt{a_p c_p-b_p^2}}{(y+\mu\mu'x')(y'-\mu\mu'x)\,i\sin
  p}\,,
  \end{eqnarray*}
  where
  \ben
  d_p=\mu\mu'(xx'-yy')+(yy'-\mu^2\mu'^2xx')\cos p.
  \ebn
  One should note that the functions $A^{+}(p)$, $\dot{A}^+(p)$ do not depend on $t$, as expected.
  \subsection{Ising model}
  In the Ising case, another parametrization is typically used.
  For simplicity, let us consider the isotropic model,
  characterized by the plaquette weight
  \ben
  W(\sigma_1,\sigma_2,\sigma_3,\sigma_4)=\exp\Bigl\{\frac12\,K(\sigma_1\sigma_2+\sigma_2\sigma_3+
  \sigma_3\sigma_4 +\sigma_4\sigma_1)\Bigr\}.
  \ebn
  Parameters of the general free-fermion model, corresponding to
  this Boltzmann weight, are given by
  \ben
  a_0=\frac{\cosh^2 K+1}{2}\,,\qquad a_4=a_{13}=a_{24}=\frac{\sinh^2 K}{\cosh^2 K+1}\,,
  \ebn\ben
  a_{12}=a_{23}=a_{34}=a_{14}=\frac{\sinh K \cosh K}{\cosh^2 K+1}\,.
  \ebn
  Let us also introduce the function $\gamma_q$, given by the positive
  root of the equation
  \ben
  \cosh\gamma_q=\sinh2K+\sinh^{-1}2K-\cos q,
  \ebn
  One can now rewrite the variables $\chi_p$ and $G_{ij}(p)$ ($i,j=1,2$) from the
  grassmann integral representation of $T^{NS}$ and $T^R$ in the
  following way:
  \ben
  \chi_p=\frac{\sinh^2 2K\,\bigl[(1+\tanh K\cos p)^2+\sin^2
  p\bigr]}{(\cosh^2K+1)^2}\,,
  \ebn
  \ben
  \chi_p\, G_{11}(p)=\chi_p\, G_{22}(p)=
  \frac{2i\sin p \,\sinh2K\,(\cosh2K-\tanh K\cos
  p)}{(\cosh^2K+1)^2}\,,
  \ebn
  \ben
  \chi_p\,G_{12}(p)=\frac{2\sinh2K}{(\cosh^2K+1)^2}\,.
  \ebn
  The eigenvalues and eigenvectors of $T^{NS}$ and $T^R$ may be found from
  \ben
  \Lambda^{NS(R)}_{max}=\left(2\sinh2K\right)^{L/2}
  \exp\left\{\frac12\;{\sum\limits_{p}}^{NS(R)}\gamma_p\right\},\qquad\qquad
  v_p/\rho_p=e^{-\gamma_p},
  \ebn
  \ben
  A^{\pm}(p)=\dot{A}^{\pm}(p)=\frac{\sinh K\cosh K\bigl[(1+\tanh K\cos p)^2+\sin^2
  p\bigr]-e^{\pm\gamma_p}}{i\sin p\;(\cosh2K-\tanh K\cos  p)}\,.
  \ebn
  It should be emphasized once again that the transfer matrix eigenvectors, constructed
  above, automatically diagonalize the translation operator $R$ as
  well. Therefore, we believe that these eigenvectors may turn out to be useful
  for the construction of a rigorous proof of the
  recently obtained formula \cite{PhysLetts} for Ising spin form factors.

  \section{Summary}
  We have obtained the transfer matrix eigenvectors of the BBS$_2$
  model on a finite lattice, using the method of grassmann
  integration. Our results are exact and explicit, i.~e. the
  eigenvectors are expressed in terms of initial lattice
  variables. Grassmann integral representation for the
  eigenvectors immediately gives their norms and allows to
  considerably advance in the computation of form factors of the
  BBS$_2$ model.

  The only two things that were actually necessary for our computation are the
  translational invariance of the model and the representation
  (\ref{grweight}) for the plaquette Boltzmann weight. In this
  respect, the method, developed in the present paper, is quite
  general and it could be extended to free-fermion models with a more
  complicated configuration space of order parameter, once these are found.

  \section*{Acknowledgements}
  The author thanks N.~Iorgov, S.~Pakuliak and V.~Shadura for
  useful discussions and communicating their results \cite{gips} prior to publication.
  Special thanks to A.~I.~Bugrij for fruitful
  collaboration. This work was supported by the grant
  INTAS--03--51--3350 and franco-ukrainian program DNIPRO.


\begin{thebibliography}{50}
  \bibitem{baxter1} R.~J.~Baxter, \textit{Superintegrable
  chiral Potts model: thermodynamic properties, an inverse model
  and a simple associated hamiltonian}, J.~Stat. Phys.~\textbf{57},
  (1989), 1--39.
  \bibitem{baxter3} R.~J.~Baxter, V.~V.~Bazhanov, J.~H.~H.~Perk,
  \textit{Functional relations for transfer matrices of the chiral
  Potts model}, Int.~J.~Mod.~Phys.~\textbf{B4}, (1990), 803--870.
  \bibitem{energy} R.~J.~Baxter, \textit{Chiral Potts model: eigenvalues of the transfer
  matrix}, Phys. Letts.~\textbf{A146}, (1990), 110--114.
  \bibitem{order} R.~J.~Baxter, \textit{The order parameter of the chiral Potts model},
  J.~Stat. Phys.~\textbf{120}, (2005), 1--36.
  \bibitem{baxter2} R.~J.~Baxter, \textit{Transfer matrix functional
  relations for the generalized $\tau_2(t_q)$ model},
  J.~Stat. Phys.~\textbf{117}, (2004), 1--25.
  \bibitem{bs} V.~V.~Bazhanov, Yu.~G.~Stroganov, \textit{Chiral Potts
  model as a descendant of the six vertex model}, J.~Stat. Phys.~\textbf{59},
  (1990), 799--817.
  \bibitem{bugITP} A.~I.~Bugrij, \textit{Partition function of the planar Ising model
  on a lattice of finite size}, preprint ITP-85-114P, Kiev, (1985).
  \bibitem{buggen} A.~I.~Bugrij, \textit{Fermionization of a generalized
  two-dimensional Ising model}, in ``Electron-electron correlation effects
  in low-dimensional conductors and superconductors'',
  eds. A.~A.~Ovchinnikov and I.~I.~Ukrainskii, Springer-Verlag,
  (1991), 135--151.
  \bibitem{bis} A.~I.~Bugrij, N.~Z.~Iorgov, V.~N.~Shadura,
  \textit{Alternative method of computation of the transfer matrix
  eigenvalues of the $\tau_2$-model for $N=2$}, JETP Letts.~\textbf{82},
  (2005), 346--351.
   \bibitem{PhysLetts} A.~I.~Bugrij, O.~Lisovyy, \textit{Spin matrix elements
   in 2D Ising model on the finite lattice},
   Phys. Letts.~\textbf{A319}, (2003), 390--394.
    \bibitem{fradkin} E.~S.~Fradkin, D.~M.~Shteingradt,
   \textit{A continuous-integral method for spin lattice models},
   Nuovo Cimento~\textbf{A47}, (1978), 115--138.
  \bibitem{gps} G.~von~Gehlen, S.~Pakuliak, S.~Sergeev,
   \textit{Bazhanov-Stroganov model from 3D approach}, J.~Phys.~\textbf{A38}, (2005),
    7269--7298.
   \bibitem{gips} G. von Gehlen, N. Iorgov, S. Pakuliak, V. Shadura,
    \textit{Baxter-Bazhanov-Stroganov model: Separation of variables and Baxter
    equation}, in preparation.
   \bibitem{kaufman} B. Kaufman, \textit{Crystal statistics. II.
   Partition function evaluated by spinor analysis}, Phys.
   Rev.~\textbf{76}, (1949), 1232--1243.
   \bibitem{smj5} M.~Sato, T.~Miwa, M.~Jimbo, \textit{Holonomic quantum fields~V},
   Publ. RIMS, Kyoto Univ.~\textbf{16}, (1980), 531--584.
   \end{thebibliography}
  \end{document}